\documentclass[12pt,a4paper]{iopart}
\usepackage{iopams}
\usepackage{graphicx}
\usepackage{setstack}
\usepackage[colorlinks=true,linkcolor=blue,citecolor=blue,filecolor=blue,urlcolor=blue]{hyperref}
\def\v{\boldsymbol v}
\def\bx{\boldsymbol x}
\def\bz{\boldsymbol z}

\def\K{\tilde{K}}
\def\ka{\tilde{\kappa}}
\def\a{\mathrm{a}}
\def\b{\mathrm{b}}
\def\c{\mathrm{c}}
\def\d{\mathrm{d}}
\def\O{\mathrm{O}}
\def\U{\mathrm{U}}
\def\V{\tilde{V}}
\def\v{\tilde{\nu}}
\def\k{\tilde{k}}

\def\fh{\hat{f}}
\def\p{\phi}
\def\ph{\hat{\phi}}

\def\Imag{\mathrm{Im}}
\def\Real{\mathrm{Re}}
\def\tD{\tau_{\mathrm{d}}}

\newcommand{\ie}{\textit{i.e.}\ }

\newcommand{\cf}{\textit{c.f.}\ }

\newcommand{\frefs}[1]{figures~\ref{#1}}
\newcommand{\srefs}[1]{sections~\ref{#1}}
\newcommand{\Frefs}[1]{Figures~\ref{#1}}

\begin{document}

\title[Transport moments beyond the leading order]{Transport moments beyond the leading order}

\author{Gregory Berkolaiko$^1$ and Jack Kuipers$^2$}
\address{$^1$ Department of Mathematics, Texas A\&M University, College Station, TX 77843-3368, USA}
\address{$^2$ Institut f\"ur Theoretische Physik, Universit\"at Regensburg, D-93040 Regensburg, Germany}
\ead{Jack.Kuipers@physik.uni-regensburg.de}

\begin{abstract}
  For chaotic cavities with scattering leads attached, transport
  properties can be approximated in terms of the classical
  trajectories which enter and exit the system.  With a semiclassical
  treatment involving fine correlations between such trajectories we
  develop a diagrammatic technique to calculate the moments of various
  transport quantities.  Namely, we find the moments of the transmission
  and reflection eigenvalues for systems with and without time
  reversal symmetry.  We also derive related quantities involving an
  energy dependence: the moments of the Wigner delay times and the density
  of states of chaotic Andreev billiards, where we find that the gap
  in the density persists when subleading corrections are included.
  Finally, we show how to adapt our techniques to non-linear
  statistics by calculating the correlation between transport moments.

  In each setting, the answer for the $n$-th moment is obtained for
  arbitrary $n$ (in the form of a moment generating function) and
  for up to the three leading orders in terms of the inverse channel
  number.  Our results suggest patterns which should hold for
  further corrections and by matching with the low order moments
  available from random matrix theory we derive likely higher order
  generating functions.
\end{abstract}


\markboth{}{}
\tableofcontents
\markboth{}{}

\section{Introduction}\label{intro}

Transport through a chaotic cavity is usually studied through a
scattering description.  For a chaotic cavity attached to two leads
with $N_1$ and $N_2$ channels respectively, the scattering matrix is
an $N\times N$ unitary matrix, where $N=N_1+N_2$.  It can be separated
into transmission and reflection subblocks
\begin{equation}
S(E) = \left(\begin{array}{cc}r_1&t' \\ t & r_2\end{array}\right),
\end{equation}
which encode the dynamics of the system and the relation between the
incoming and outgoing wavefunctions in the leads.  The unitarity of the 
scattering matrix $S^{\dagger}S=I=SS^{\dagger}$
leads to the following relations (among others)
\begin{equation} \label{refltransrel}
r_{1}^{\dagger}r_{1}+t^{\dagger}t = I_{N_{1}}, \qquad r_{2}r_{2}^{\dagger}+tt^{\dagger} = I_{N_{2}},
\end{equation}
while the transport statistics themselves are related to the terms 
in \eref{refltransrel} involving the scattering matrix (or its transmitting 
and reflecting subblocks) and their transpose conjugate.  For example, the 
conductance is proportional to the trace $\Tr \left[t^\dagger t\right]$ 
(Landauer-B\"uttiker formula \cite{Landauer57,Buttiker86,Landauer88}), while other physical
properties are expressible through higher moments like $\Tr \left[t^\dagger t\right]^n$.

There are two main approaches to studying the transport statistics in clean ballistic systems:
a random matrix theory (RMT) approach, which argues that $S$ can be viewed as a
random matrix from a suitable ensemble, and a semiclassical approach
that approximates elements of the matrix $S$ by sums over open scattering
trajectories through the cavity.

It was shown by Bl\"umel and Smilansky
\cite{BluSmi_prl88,BluSmi_prl90} that the scattering matrix of a
chaotic cavity is well modelled by the Dyson's circular ensemble of
random matrices of suitable symmetry.  Thus, transport properties of
chaotic cavities are often treated by replacing the scattering matrix
with a random one (see \cite{beenakker97} for a review).  The
eigenvalues of the transmission matrix $t^{\dagger}t$ then follow a
joint probability distribution, which depends on whether the system
has time reversal symmetry or not, and from which transport moments
and other quantities can be derived.  Though the conductance and its
variance were known for arbitrary channel number \cite{bm94,jpb94},
other quantities were limited to a diagrammatic expansions in inverse
channel number, see \cite{bb96}.  However, the RMT treatment has
recently experienced a resurgence due to the connection to the Selberg
integral noticed in \cite{ss06}.  Following the semiclassical result
for the shot noise \cite{braunetal06}, the authors of \cite{ss06} used
recursion relations derived from the Selberg integral to calculate the
shot noise and then later all the various moments up to fourth order
for arbitrary channel number \cite{ssw08}.

Since then a range of transport quantities have been treated, for
example the moments of the transmission eigenvalues for chaotic
systems without time reversal symmetry (the unitary random matrix
ensemble) \cite{vv08,novaes08,sm10,lv11} and those with time reversal symmetry (the
orthogonal random matrix ensemble) \cite{sm10,lv11}.  For the unitary ensemble, the moments of the
conductance itself were also obtained in \cite{novaes08} and, using a 
different approach, in \cite{ok08} which was later extended to the 
moments of the shot noise \cite{ok09}.  Building again on the Selberg integral approach,
the moments of the conductance and shot noise have been derived for both 
symmetry classes \cite{kss09}.  Interestingly these results,
though all exact for arbitrary channel number, are given by different
combinatorial sums, and the question of how they are related to each
other is still open in many cases.

On the other hand, the semiclassical approach makes use of the following
approximation for the scattering matrix elements
\cite{miller75,richter00,rs02}
\begin{equation} 
  \label{scatmateqn}
  S_{oi}(E) \approx \frac{1}{\sqrt{N\tD}}\sum_{\gamma (i \to o)}
  A_{\gamma}(E)\rme^{\frac{\rmi}{\hbar}S_{\gamma}(E)} ,
\end{equation}
which involves the open trajectories $\gamma$ which start in channel
$i$ and end in channel $o$, with their action $S_{\gamma}$ and
stability amplitude $A_{\gamma}$.  The prefactor also involves $\tD$
which is the average dwell time, or time the trajectory spends inside
the cavity.  For transport moments we consider quantities of the type
\begin{equation}
\label{semitrajeqn}
\fl \left\langle\Tr \left[X^{\dagger}X\right]^{n} \right\rangle \sim \left\langle \frac{1}{{(N\tD)}^{n}}
 \prod_{j=1}^{n} \sum_{{i_j,o_j}} 
  \sum_{\substack{\gamma_j(i_j\to o_j) \cr 
      \gamma'_j (i_{j+1}\to o_{j})}} A_{\gamma_j}A_{\gamma'_j}^{*}
  \rme^{\frac{\rmi}{\hbar}(S_{\gamma_j}-S_{\gamma'_j})} \right\rangle,
\end{equation}
where the trace means we identify $i_{n+1}=i_1$ and where $X$ is
either the transmitting or the reflecting subblock of the scattering
matrix.  The averaging is performed over a window of energies $E$.

The choice of the subblock $X$ affects the sums over the possible
incoming and outgoing channels, but not the trajectory structure which
involves $2n$ classical trajectories connecting channels.  Of these,
$n$ trajectories $\gamma_j$, $j=1,\ldots,n$, contribute with positive
action while $n$ trajectories $\gamma'_j$ contribute with negative
action.  In the semiclassical limit of $\hbar\to0$ we require that
these sums cancel on the scale of $\hbar$ so that the corresponding
trajectories can contribute consistently when we apply the averaging
in \eref{semitrajeqn}.

\begin{figure}
\center
\includegraphics[width=\textwidth]{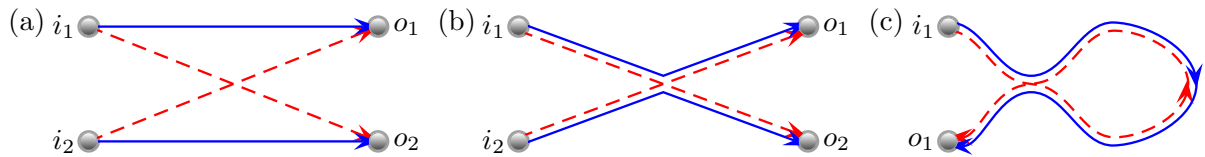}
\caption{\label{firstdiag}(a) The semiclassical trajectories for the 
second moment travel around a closed cycle.  By collapsing the trajectories 
onto each other as in (b) we can create a small action difference and 
a trajectory quadruplet with a single encounter that contributes to the 
shot noise at leading order in inverse channel number.  (c) For the conductance, 
a trajectory pair with a similar encounter provides the first subleading 
order correction for systems with time reversal symmetry.}
\end{figure}

The main idea of the semiclassical treatment is that, in order to
achieve a small action difference, the trajectories $\{\gamma'_j\}$,
must follow the path of trajectories $\{\gamma_j\}$ most of the time,
deviating only in small regions called \emph{encounters}.  This is
best illustrated with an example.  In \fref{firstdiag}(a) the
schematic depiction of the trajectories is shown for the case $n=2$.
We have 2 trajectories $\gamma$ depicted by solid lines,
$\gamma_1:i_1\to o_1$ and $\gamma_2:i_2\to o_2$, and 2 trajectories
$\gamma'$ depicted by dashed lines, $\gamma'_1: i_2\to o_1$ and
$\gamma'_2: i_1 \to o_2$.  \Fref{firstdiag}(b) shows one
possible configuration for achieving a small action difference:
trajectory $\gamma'_1$ departs from the incoming channel $i_2$
following the path of trajectory $\gamma_2$.  Then, when the
trajectories $\gamma_1$ and $\gamma_2$ come close to each other in
phase space (thus the term `encounter'), trajectory $\gamma'_1$
switches from following $\gamma_2$ to following $\gamma_1$, before
arriving at its destination channel $o_1$.  The trajectory $\gamma'_2$
does the opposite.  The picture in \fref{firstdiag}(b) is
referred to as a `diagram'; it describes the topological configuration
of the trajectories in question, while leaving out metric details.
The task of semiclassical evaluation can therefore be divided into two parts:
evaluation of the contribution of a given diagram by integrating over
all possible trajectories of given structure and enumeration of all
possible diagrams.

Historically, the semiclassical treatment started with the mean
conductance $\left\langle\Tr \left[t^{\dagger}t\right]\right\rangle$,
involving a single trajectory and its partner.  The leading order
contribution comes from trajectory pairs that are identical --- the
so-called diagonal approximation which was evaluated in
\cite{bjs93a,bjs93b}.  The first non-diagonal pair was treated in
\cite{rs02} and involved a single encounter where one trajectory had a
self-crossing while the partner avoided crossing as in
\fref{firstdiag}(c).  Such a pair can only exist when the system has
time reversal symmetry and its contribution was shown to be one order
higher in inverse channel number, $1/N$, than the diagonal terms.  The
expansion to all orders in inverse channel number was then performed
in \cite{heusleretal06} for systems with and without time reversal
symmetry by considering arbitrarily many encounters each involving
arbitrarily many trajectory stretches.

Importantly, the work on the conductance \cite{heusleretal06} showed
that the semiclassical contribution of a diagram can be decomposed
into a product over the constituent parts of the diagram, greatly
simplifying the resulting sums.  In fact for the second moment, the
shot noise, all such diagrams were generated in \cite{braunetal06} and
with them the full expansion in inverse channel number.  This, along
with the conductance variance and other transport correlation
functions, as well as the semiclassical background was covered in
detail in \cite{mulleretal07}.

However, the method for diagram enumeration considered in
\cite{heusleretal06,braunetal06,mulleretal07} becomes unwieldy for
higher moments, which encode finer transport statistics.  To the 
leading order in $1/N$, the higher moments were derived in \cite{bhn08}.
The semiclassical approach requires a large number of channels in each lead, 
$1\ll N_1,N_2$, but to unambiguously separate the orders in inverse 
channel number one may additionally assume that both $N_1$ and $N_2$ 
are of the same order as $N$.  For example, the result of \cite{bhn08} 
was in terms of the variable $\xi=N_1 N_2/N^2$ which should then be constant
and introduce no further channel number scaling.  We therefore make the  
same assumption in this article when describing the different orders in $1/N$, though
of course a different scaling, say keeping $N_1$ fixed so that 
$\xi\sim\mathrm{O}(1/N)$, may simply lead to a mixing of the different `orders' 
without changing the individual results.

The diagrams contributing at the leading order to the $n$-th moment
were shown in \cite{bhn08} to be trees.  The tree expansions turned
out to be very well suited to analysis of other interesting physical
quantities, such as the statistics of the Wigner delay times
\cite{bk10}, which are a measure of the time spent in the scattering
region, and the density of states in Andreev billiards
\cite{kuipersetal10,kuipersetal10b}.  If we imagine replacing the
scattering leads by a superconductor we have a closed system called an
Andreev billiard.  Each time an electron inside the system hits the
superconductor it is reflected as a hole retracing its path until it
hits the superconductor and is retroreflected as an electron again.
Wave interference between these paths leads to significant effects,
most notably a complete suppression of the density of states for a
range of energies around the Fermi energy.  Similarly, strong effects
on the conductance (of the order of the mean conductance) can also be
seen if we attach additional superconducting leads to our original
chaotic cavity (making a so-called Andreev dot) \cite{wj09,ekr10}.
The size of these effects make such systems especially interesting for
a semiclassical treatment.  But treating these effects effectively
requires knowledge of \emph{all} the higher moments, and gives us strong
reason to go beyond low $n$.

One particular nicety of the semiclassical approach is that it can incorporate, 
in a natural way, the effect of the Ehrenfest time.  This is the time scale that 
governs the transition from classical dynamics to wave interference, which dominates when 
the Ehrenfest time is small (on the scale of the typical dwell time).  
For larger Ehrenfest times, the competition between the different types of 
behaviour leads to quite striking features, like an additional gap both in 
the density of states of Andreev billiards and in the probability distribution of 
the Wigner delay times \cite{kuipersetal10b,wkr10}.  Semiclassically we can explicitly track the 
effect of the Ehrenfest time all the way to the `classical' limit, which can only be 
achieved using RMT by postulating the Ehrenfest time dependence of the scattering 
matrix.

Alongside the case of ballistic systems, the typical chaotic behaviour
and transport statistics can also be induced by introducing disorder in the system.
For weak disorder the transport properties coincide with those obtained from 
RMT, and one can also obtain the full counting statistics at leading order, 
as well as weak localisation corrections and universal
conductance fluctuations, using circuit theory \cite{nb09}.
If the disorder averaging is treated using diagrammatic perturbation theory (see, for
example, \cite{am07}) multiple scattering events can be summed, in the limit of weak
disorder, as a ladder diagram known as a Diffuson.  This corresponds to
the parts of semiclassical diagrams where the trajectories are nearly identical, 
as on the left of the encounter in \fref{firstdiag}(c).  The
disordered systems' counterpart of the loop on the right of 
the encounter in \fref{firstdiag}(c), traversed by trajectories in opposite
directions, is called a Cooperon,
while the encounter itself corresponds to a Hikami box \cite{hikami81}.  Although
transport properties like the weak localisation diagram related to
\fref{firstdiag}(c) and conductance and energy level fluctuations \cite{as86} can be treated 
diagrammatically, usually powerful field-theoretic methods involving the 
nonlinear $\sigma$ model are used (see \cite{lerner03} for an introduction).  
These methods can treat both weak and stronger disorder non-perturbatively, and by using
supersymmetry \cite{efetov82,efetov97} a large range of transport and
spectral properties can be obtained, for open and closed systems correspondingly.
More importantly, the applicability of RMT for weakly disordered systems
can be justified and RMT shown to be the zero-dimensional variant of
the $\sigma$ model \cite{efetov83,vwz85}.  Alongside the supersymmetric $\sigma$ model,
there is also the replica $\sigma$ model which is particularly useful
for perturbative expansions. This leads to a diagrammatic expansion,
with diagrams that can be reinterpreted
as correlated semiclassical trajectories \cite{sla98}.  In fact this connection between 
semiclassical diagrams and disorder diagrams from the replica $\sigma$ model lay behind
the semiclassical treatment of energy level correlations in closed systems 
\cite{sr01,mulleretal04,mulleretal05} which in turn led to the semiclassical treatment
of transport \cite{rs02,heusleretal06,mulleretal07} discussed above.

To summarize, there are established semiclassical tools for the analysis of
\eref{semitrajeqn} for small $n$ to all orders of $1/N$ and for all
$n$ but only to the leading order of $1/N$.  It is the purpose of this
article to start closing this gap.  For all $n$ we derive the next
two corrections for \eref{semitrajeqn} and related quantities.  We
show that the contributing diagrams can be generated by grafting trees
onto the `base diagrams', which can be obtained by `cleaning' the
diagrams used in \cite{heusleretal06}.  We therefore first 
review the leading order tree recursions in \sref{subtrees} before 
treating transport moments beyond the leading order in \sref{transportmoments}.
We start by cleaning the diagram of \fref{firstdiag}(c) which gives the first subleading 
order orthogonal correction.  Grafting trees onto the 
base diagram leads to a generating function, which we apply 
to calculate the moments of the transmission and reflection eigenvalues.  
Proceeding to the next order in $1/N$ we then treat the second 
subleading order diagrams for the unitary and  
the orthogonal case.  For the moments of the reflection 
and transmission eigenvalues we find that our generating functions 
simplify and become rather straightforward.

The graphical recursions we use provide a new insight into the leading 
order terms which is particularly 
useful for energy dependent correlation functions.  Such correlation functions 
are needed for a treatment of the density of states of Andreev billiards, which we 
consider in \sref{andreev} where we find that the hard gap, previously found at 
leading order in $1/N$, persists at least for the next two orders.  Also derivable 
from energy dependent correlation functions are the moments of the Wigner delay 
times, treated in \sref{wigner}, and we find that the corrections at each order in 
$1/N$ are also generated by relatively simple functions.  Of course, the transport 
moments in \eref{semitrajeqn} are only one type of transport quantity, and we finally
look at non-linear statistics in \sref{twotraces} and see how 
their treatment follows naturally from the previous semiclassical considerations.

We shall be comparing our semiclassical results with the prediction 
of RMT, where those predictions are available: previously 
(of the quantities treated here) only the moments of the transmission 
amplitudes for systems without time reversal symmetry have been given 
for an arbitrary number of channels \cite{vv08,novaes08}.  Explicit 
results for systems with time reversal symmetry have just been derived \cite{sm10,lv11} 
and we were pleased that \cite{sm10} shared those results with us beforehand.  
The moments of the Wigner delay times for both symmetry classes have also 
been obtained \cite{sm10}.

Of the recent RMT results, it is those concerned with the asymptotic 
expansion as the number of channels increases, currently to leading order 
\cite{novaes07,carreetal10,krattenhaler10}, 
that particularly connect with the work here.  Semiclassically, without 
the equivalent of the Selberg integral, we are still restricted to an 
expansion in inverse powers of the channel number, but as we shall 
see the semiclassical treatment leads to explicit and surprisingly 
simple generating functions at each order in inverse channel number. 
This simplicity until now remained hidden in the combinatorial sums of 
the RMT results and may suggest ways of simplifying those results and 
of highlighting the underlying combinatorial structure.

\addtocontents{toc}{\vspace{-1em}}
\section{Subtrees}\label{subtrees}

The semiclassical treatment of the conductance beyond the diagonal contribution, 
starting \cite{rs02} with the trajectory pair depicted in \fref{firstdiag}(c), 
required two main ingredients.  The first was to estimate how often a trajectory 
would come close to itself and have a self-encounter.  This is performed using 
the global ergodicity of the chaotic dynamics.  The second was that, given such an 
encounter, we can use the local hyperbolicity of the motion to find the partner 
trajectory which reconnects the stretches of the original trajectory
in a different way.  Then one can determine 
the action difference between the two trajectories and hence their contribution in 
the semiclassical limit.  When treating diagrams with more numerous and more complicated 
encounters, the authors of \cite{heusleretal06} showed that these two ingredients allowed them to 
express the total contribution as a product of integrals over the encounters and over 
the `links', the trajectory stretches which connect the encounters together.  Performing 
these integrals then led to simple rules for the contributions of the constituent parts 
of any diagram, and essentially reduced the problem down to the combinatorial one of 
finding all the possible diagrams.  For the first two transport moments, this was 
done \cite{mulleretal07} by cutting open the periodic orbit pairs that contribute to 
spectral statistics \cite{sr01,mulleretal04,mulleretal05}.

For the higher moments, as shown in \cite{bhn08}, the diagrams that
contribute at leading order in inverse channel number are rooted plane
trees.  The reason is simple: according to the semiclassical
evaluation rules of \cite{mulleretal07}, every encounter contributes
a factor of $-N$ while every link contributes a factor of $1/N$.
The leading order is thus achieved by a diagram with the minimal
possible difference between the number of links (edges) and encounters
(internal vertices).  It is a basic fact of graph theory that this
difference is minimized by trees; each independent cycle in a graph
adds one to this difference.  Thus to go beyond the leading order one
needs to consider diagrams with an increasing number of cycles.  We
will approach this task by describing the topology of the cycles using
`base diagrams' --- graphs with no vertices of degree 1 or 2 --- and
then grafting subtrees onto the base diagrams.

Adding a subtree does not change the order of the contribution in
inverse channel number $1/N$ but adds more incoming and outgoing
channels thus changing the order of the moment $n$.  Because
we will be joining the trees to existing structures, unlike the
treatment in \cite{bhn08,bk10,kuipersetal10,kuipersetal10b}, here we
do not root our trees in an incoming channel, but at an arbitrary
point.  These trees then correspond to the restricted trees in
\cite{bhn08,kuipersetal10b} and will be referred to as
`subtrees'.  We also note that the generating function variables we
use here have slightly different definitions than in
\cite{bhn08,bk10,kuipersetal10,kuipersetal10b}.  Our present choice is
more appropriate for the subleading orders and the different transport
quantities considered.

We now summarize the derivation of the subtree generating functions
which were introduced in \cite{bhn08} and further developed in
\cite{bk10,kuipersetal10b}.  A subtree consists of a root, several
vertices of even degree (called `nodes', they correspond to encounters
between various trajectories) and $2n-1$ vertices of degree one
(called `leaves', they correspond to incoming or outgoing channels).  The
leaves are labelled $i$ or $o$ alternatingly as we go around the tree
anti-clockwise.  There are two types of subtrees: the $f$-subtrees
have leaves labelled 
$o_k$, $i_{k+1}$, $o_{k+1}$, $i_{k+2}$ etc.  The label $i_k$
would correspond to the root if we were to label it too.  The
$\fh$-subtrees have leaf labels $i_{k+1}$, $o_{k+1}$, $i_{k+2}$,
$o_{k+2}$ etc.  The reference index $k$ depends on the location of the
subtree on the diagram.

It is possible that an encounter happens immediately as several
trajectories enter the cavity from the lead or exit the cavity into the lead.
To keep account of these situations, we say that an $l$-encounter (node of degree $2l$) may
`$i$-touch' the lead if it is connected directly to $l$ incoming
channels (leaves with label $i$) and `$o$-touch' if connected to $l$
outgoing channels.  When an encounter touches the lead, the edges
connecting it to the lead get cut off and all the channels must coincide,
although in the diagrams we keep short `stubs' to avoid changing the 
degree of the encounter vertex.

\begin{figure}
\center
\includegraphics[width=0.5\textwidth]{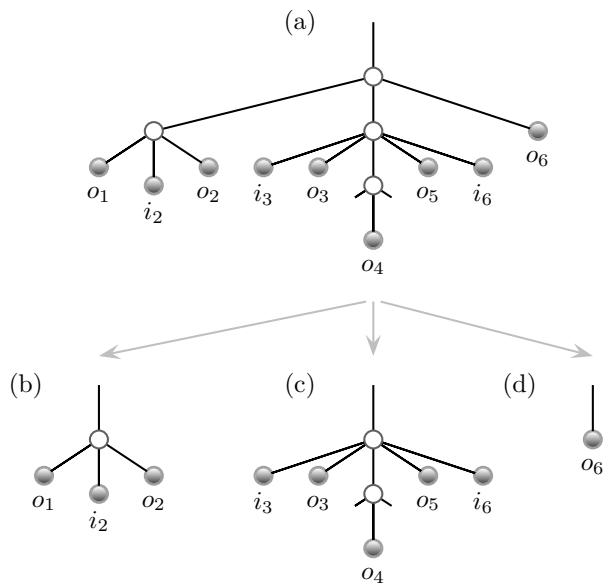}
\caption{\label{treecutting}The subtree shown in (a) is cut at its 
top node (of degree 4) creating subtrees (b)-(d).  Subtree (c) has 
the incoming and outgoing directions reversed.  The lower vertex in (a), 
and hence (c), is $i$-touching the lead so that the channels $i_4$ and $i_5$ 
(not shown) coincide ($i_4=i_5$).  This is represented by the short 
stubs, and the encounter now starts in the incoming lead.}
\end{figure}

We define the generating functions $f(\bx, \bz_{i}, \bz_{o})$ and $\fh
= f(\bx, \bz_{o}, \bz_{i})$ which are counting $f$- and $\fh$-trees
correspondingly.  The meaning of the variables $\bx=(x_2,x_3,\ldots)$,
$\bz_i=(z_{i,2}, z_{i,3},\ldots)$ and $\bz_o=(z_{o,2},
z_{o,3},\ldots)$ is as follows:
\begin{itemize}
\item $x_l$ enumerate the $l$-encounters that do not touch the lead,
\item $z_{i, l}$ enumerate the $l$-encounters
  that $i$-touch the lead,
\item $z_{o, l}$ enumerate the $l$-encounters
  that $o$-touch the lead.
\end{itemize}
For example, the coefficient of $x_3x_2^2z_{i,2}$ gives the number of
trees with one 3-encounter (a vertex of degree 6) and three
2-encounters (vertices of degree $4$), one of which $i$-touches the
lead.  An example of such a tree is given in
\fref{treecutting}(a).  We note that if an encounter may touch the
lead, the generating function includes (and sums) both possibilities: touching
and non-touching.  For example, the left-most vertex of the tree in
\fref{treecutting}(a) may $o$-touch the lead, but this possibility
is counted separately.

In addition we will use several secondary parameters that will allow
us to adapt the subtree generating functions to each of the four
quantities considered in the paper.  These parameters are:
\begin{itemize}
\item $y$ is the semiclassical contribution of an edge (link)
\item $c_i$ and $c_o$ are the contributions of an incoming and
  an outgoing channel
\item $\sigma$ is a special correction parameter for the situation
  when an $o$-touching node is directly connected to an $i$-channel
  ($\sigma=0$ everywhere apart from \sref{wigner}).
\end{itemize}

We obtain a recursion for the functions $f$ and $\fh$ by cutting the
subtree at the top encounter node.  If this node is of degree $2l$,
this leads to $2l-1$ further subtrees as illustrated in
\fref{treecutting}.  Assuming we started with an $f$-subtree, $l$ of
the new subtrees also have type $f$, while the remaining $l-1$ are
$\fh$-subtrees.  Thus an $f$-subtree with an $l$-encounter at the top
contributes $yx_l f^l \fh^{l-1}$ to the generating function $f$.
Additionally, we consider the possibility for the top node of an
$f$-subtree to $o$-touch.  In this case its odd-numbered further
subtrees are empty stubs and the even-numbered subtrees are still arbitrary,
leading to the contribution $yz_{o,l}(\fh+\sigma)^{l-1}$.  Here we
have included a correction term $\sigma$ which is used in
\sref{wigner} to control the contribution of any $\fh$-subtree that
consists of one edge and directly connects an incoming and outgoing
channel and is set to 0 in the rest of the paper.

We start our recursion relation at the value for an empty tree, which
consists of a link (with the factor $y$) and an outgoing channel
(providing a factor $c_{o}$),
\begin{equation}
  \label{eq:recur_f}
  f = yc_{o} + y\sum_{l=2}^\infty \left[ x_l f^{l}
    {\fh}^{l-1} + z_{o,l} (\fh+\sigma)^{l-1} \right].
\end{equation}
The recursion is similar for $\fh$, with the roles of $i$- and
$o$-variables switched,
\begin{equation}
  \label{eq:recur_fhat}
  \fh = yc_{i} + y\sum_{l=2}^\infty \left[ x_l {\fh}^{l}
    {f}^{l-1} + z_{i,l} f^{l-1} \right] .
\end{equation}

\subsection{Reflection} \label{subtreesreflection}

For the reflection into lead 1 we will consider the generating function 
\begin{equation}
R(s) = \sum_{n=1}^{\infty} s^n \left\langle\Tr [r_{1}^{\dagger}r_{1}]^n\right\rangle ,
\end{equation}
where the power of $s$ counts the order of the moments.  For the individual semiclassical
diagrams we make use of the diagrammatic rules of \cite{mulleretal07}, where each link 
contributes a factor of $1/N$ while each encounter provides the factor $-N$.  Each channel 
is in lead 1, so can be chosen from the $N_1$ available and provides this factor.  When
an encounter starts (or ends) in the lead, all the incoming (or outgoing) channels must then
coincide in the same channel, leading again to the factor $N_1$.  Bearing in mind the
meaning of the variables introduced above, we therefore have to make the following semiclassical
substitutions:
\begin{equation}
\label{reflsemivalues}
\fl y=\frac{1}{N},\quad x_{l}=-N, \quad z_{i,l}=z_{o,l}=r^lN_{1},\quad c_{i}=c_{o}=rN_{1},\quad \sigma=0,
\end{equation}
where we have introduced $r$ whose power counts the total number 
of channels and which allows us to keep track of the total 
contributions to different moments.  The $n$-th moment
involves $2n$ channels so we have the relation $s=r^2$.   
Each channel factor $c$ then includes the factor $r$,
while the formula for $z_{i,l}$ in \eref{reflsemivalues} 
accounts for the fact that when an $l$-encounter enters the 
incoming channels we have $l$ channels coinciding 
but only a single channel factor.

If we define $\zeta_{1}=N_{1}/N$, the subtree recursions \eref{eq:recur_f} and \eref{eq:recur_fhat} both become
\begin{equation}
  \label{frefleqn}
  f = r\zeta_{1} - \sum_{l=2}^\infty f^{2l-1} + r\zeta_{1}\sum_{l=2}^\infty r^{l-1}f^{l-1} .
\end{equation}
Performing the sums (where the terms $f$ and $r\zeta_1$ correspond to $l=1$ of the sums) this is
\begin{equation}
  \label{freflsummed}
  0=-\frac{f}{1-f^2}+\frac{r\zeta_{1}}{1-rf} ,
\end{equation}
which can be written as the quadratic
\begin{equation}
  \label{freflquad}
  r(1-\zeta_{1})f^2-f+r\zeta_{1} =0, \qquad f=\frac{1-\sqrt{1-4\xi r^2}}{2r(1-\zeta_{1})} ,
\end{equation}
where $\xi=\zeta_{1}(1-\zeta_{1})$ and where we take the solution whose expansion 
agrees with the contributions of the semiclassical diagrams.

\subsection{Transmission} \label{subtreestransmission}

For the transmission we treat the function
\begin{equation}
T(s) = \sum_{n=1}^{\infty} s^n \left\langle\Tr [t^{\dagger}t]^n\right\rangle,
\end{equation}
and to distinguish it more clearly from the reflection we will 
call the corresponding subtree generating function 
$f=\p$ here.  For the transmission, the equations are 
a bit more complicated than for the reflection because 
$\p\neq\ph$ in general.  For the substitution we need
\begin{eqnarray}
y&=&\frac{1}{N},\quad x_{l}=-N, \quad z_{i,l}=r^lN_{1},\quad z_{o,l}=r^lN_{2},\nonumber \\
c_{i}&=&rN_{1},\quad c_{o}=rN_{2},\quad \sigma=0, \label{transsemivalues}
\end{eqnarray}
where the only difference from \eref{reflsemivalues} is that the outgoing channels are
now in lead 2 and can be chosen from the $N_2$ available.

The contribution of the subtrees \eref{eq:recur_f} once summed becomes
\begin{equation}
  \label{ftranssummed}
  0=-\frac{\p}{1-\p\ph}+\frac{r\zeta_{2}}{1-r\ph}, 
\qquad\mbox{or}\qquad \p=r\zeta_{2} + r\zeta_{1}\p\ph ,
\end{equation}
with $\zeta_{2}=N_{2}/N$ and using that $\zeta_1+\zeta_2=1$.  
Likewise \eref{eq:recur_fhat} becomes
\begin{equation}
  \label{fhattranssummed}
  0=-\frac{\ph}{1-f\ph}+\frac{r\zeta_{1}}{1-r\p}, 
\qquad\mbox{or}\qquad \ph=r\zeta_{1} + r\zeta_{2}\p\ph,
\end{equation}
where, as before $\xi = \zeta_1(1-\zeta_1) = \zeta_1\zeta_2$.
With $h=\p\ph$ we have
\begin{equation}
\label{htranseqn}
r^2\xi h^2 + [r^2(1-2\xi)-1]h +r^2\xi=0 ,
\end{equation}
from which we can find the equations
\begin{equation}
\label{ftransfinaleqn}
\fl \p^2 -\left(\frac{1-r^2}{r\zeta_{2}}+2r\right)\p +1=0, 
\qquad \ph^2 -\left(\frac{1-r^2}{r\zeta_{1}}+2r\right)\ph +1=0 .
\end{equation}

\addtocontents{toc}{\vspace{-1em}}
\section{Transport moments} \label{transportmoments}

By a simple counting argument, the order of a diagram in terms of
inverse channel number is the number of edges minus the number of
vertices (both leaves and nodes).  Thus a diagram contributes at the
order $(1/N)^{\beta-1}$, where $\beta$ is the number of the
independent cycles in the diagram (also known as the cyclomatic number
or the first Betti number, hence the notation).  The leading
contribution thus comes from tree diagrams which have $\beta=0$ and
the next contribution comes from diagrams with one cycle.

\subsection{First orthogonal correction} \label{orthogonal}

\begin{figure}
\center
\includegraphics[width=\textwidth]{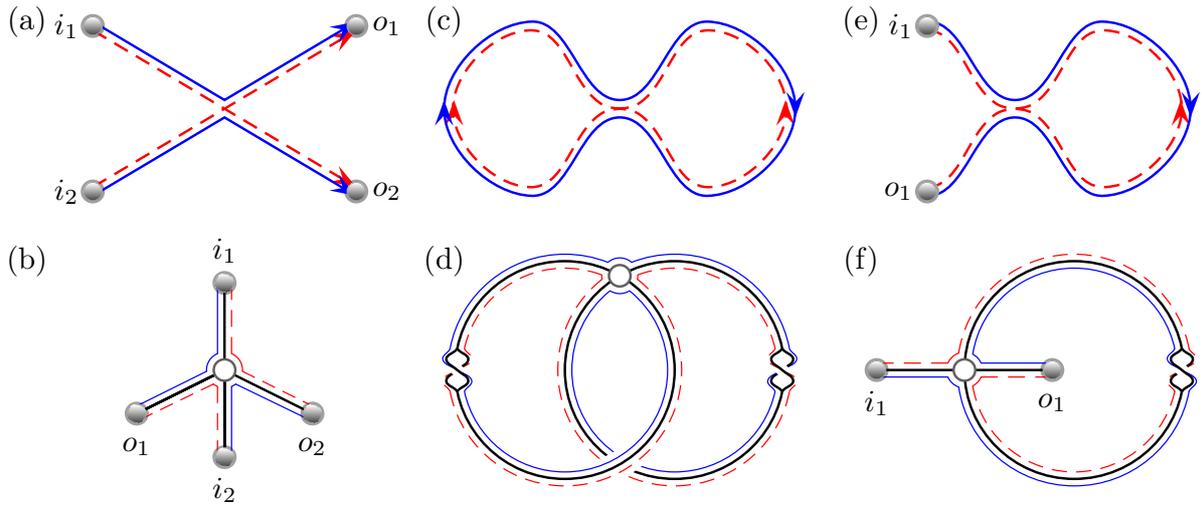}
\caption{\label{boundarywalks}The correlated trajectory quadruplet 
in (a) which contributes to the second moment at leading order 
in inverse channel number can be redrawn as the ribbon tree in (b) 
by `untwisting' the encounter.  The four trajectories themselves 
can be read off from the boundary walk shown.  At subleading order 
in inverse channel number, we start with the correlated periodic 
orbit pair in (c) which can be represented as the graph in (d) 
with corresponding boundary walks.  Cutting the periodic orbit 
along the left link (which is traversed in the same direction 
by the orbit and its partner) creates the correlated trajectory 
pair in (e) which contributes to the first moment.  Changing this 
diagram into a graph we arrive at the structure in (f) which is 
a M\"obius strip with an empty subtree inside and outside the loop.
The intertwined S's in diagrams (d) and (f) represent twists in the
corresponding ribbon links.}
\end{figure}

A diagram with one cycle can be thought of as a loop with trees grafted
on it.  But there is a twist.  The reconstruction of the trajectories'
structure from a tree, see \cite{bhn08}, was done by means of the
boundary walk.  It helps to visualise the edges of the tree as strips,
a model that is called a \emph{fat} or \emph{ribbon} graph in
combinatorics.  This fixes the circular order of edges around each
vertex and, going along the boundary, prescribes a unique way to
continue the walk around a vertex (see \cite{Zvo_mcm97} for an
accessible introduction).  The trajectories $\gamma_j$ of
equation~\eref{semitrajeqn} are then read as the portions of the walk
going from $i_j$ to $o_j$.  The trajectories $\gamma_j'$, on the other
hand, appear in reverse as portions of the walk going from $o_j$ to
$i_{j+1}$.  For example, the diagram in \fref{boundarywalks}(a) which contributes 
at leading order in $1/N$ can be redrawn as the tree in \fref{boundarywalks}(b) with 
the corresponding boundary walk shown.

The trace in \eref{semitrajeqn} means that the boundary walk is
closed and the equality of total actions implies that each edge of the
diagram is traversed twice (once by $\gamma$ and once by $\gamma'$).
This means that a valid diagram must have \emph{one face}.  In
particular, there must be a way for the walk to cross from inside to
outside of the cycle of the first correction diagram.  The diagram
thus has the topology of a M\"obius strip with (ribbon) trees grafted
on the edges.  We will refer to the diagram without any trees
(M\"obius strip in this case) as the \emph{base diagram} or
\emph{structure}.

\begin{figure}
\center
\includegraphics[width=\textwidth]{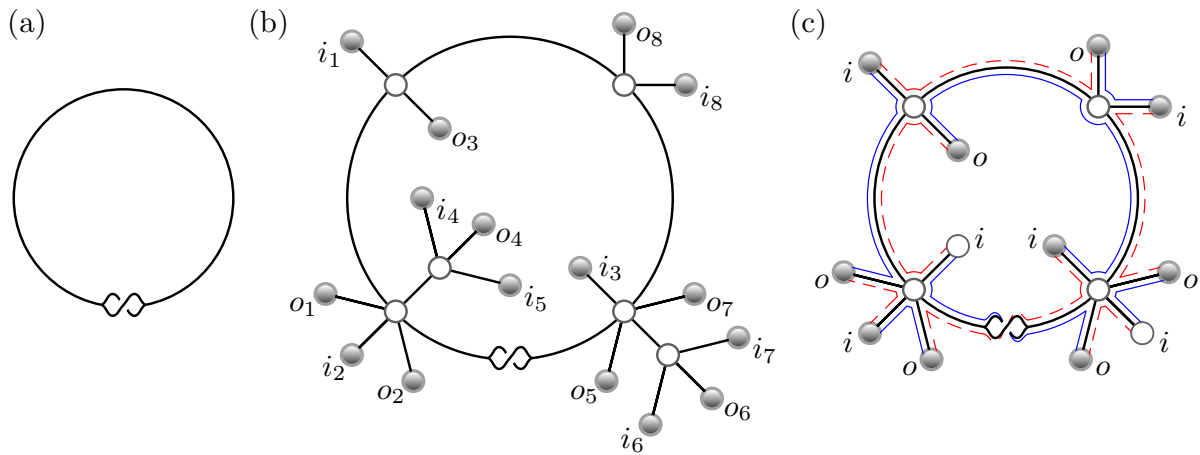}
\caption{\label{orthotree}To obtain the base structure in (a) we 
can simply remove the empty subtrees of \fref{boundarywalks}(f).  
Appending subtrees to (a) we can then create all the possible graphs, 
but for the graph to remain a M\"obius strip we need an odd number of odd nodes, 
as for example in the graph in (b).  We draw the boundary walk in (c) 
where we truncated the subtrees at their first node as they always 
have an odd number of leaves thereafter.  The top left and bottom 
right nodes along the M\"obius strip in (b) or (c) may also enter 
the lead for reflection quantities.}
\end{figure}

It is also beneficial to consult the full expansion in powers of the
inverse channel number of the first two moments of the transmission
eigenvalues \cite{rs02,heusleretal06,braunetal06,mulleretal07} and
to draw the corresponding diagrams as ribbon graphs.
The procedure of going from the closed periodic orbits to scattering
trajectories and then to a graph is illustrated in
\fref{boundarywalks} for the first subleading order correction.  Removing 
the remaining subtrees from \fref{boundarywalks}(f) leads to the base structure 
in \fref{orthotree}(a) to which we can append subtrees to create valid diagrams 
like \fref{orthotree}(b) whose boundary walk is depicted in \fref{orthotree}(c).  
As the base structure involves a loop which is traversed in opposite 
directions by the trajectory and its partner, all the diagrams created in 
this way can only exist in systems with time reversal symmetry 
(corresponding to the orthogonal RMT ensemble).

Along the loop we can add subtrees at any point and to make a valid
$l$-encounter we must add $2l-2$ subtrees (the remaining two stretches
in the encounter belong to the loop itself).  If the node has an odd
number of trees both inside and outside the loop, we refer to it as an
\emph{odd node}.  It is easy to convince oneself that in order to have
each stretch of the loop traversed once by a $\gamma$-trajectory and
once by a $\gamma'$-trajectory there must be an odd number of odd
nodes around the loop.

We start by evaluating the contribution of a node along the loop.  For
the node we include all possible sizes $l$ of the resulting encounter.
Adding the $2l-2$ subtrees (of which $l-1$ start with an incoming
direction and $l-1$ with an outgoing direction) there are $2l-1$ ways
of splitting them into groups inside or outside of the encounter.
With the $l-1$ ways which result in an odd node we include the factor
$p$ whose power will later count the total number of odd nodes around
the loop.  This leads to
\begin{equation}
\label{Aeqn}
A(p)=\sum_{l=2}^{\infty}x_{l}(f\fh)^{l-1}[p(l-1)+l] .
\end{equation}

The number of incoming and outgoing channels connected to the same node is
equal [see an example in \fref{orthotree}(b)].
An $l$-encounter can touch the lead only if every \emph{other} edge connected
to it is empty (connected directly to a leaf).  Since for nodes on the loop we need to include the
edges that belong to the loop itself and which cannot be empty, we conclude
that only odd nodes can possibly touch the lead.  Since in this case
we need $l$ empty edges and we have $l-1$ edges of each type, touching
the lead is possible only if the incoming and outgoing channels are in
the same lead, as they are when we consider a reflection quantity.

With $2k-1$ trees on the inside of which $k$ must be empty (and the
remaining $k-1$ arbitrary) and the remaining $2l-2k-1$ on the outside
(with $l-k$ empty and $l-k-1$ arbitrary) and with
$\bz_{i}=\bz_{o}=\bz$ for reflection quantities we add the following
to the node contribution
\begin{equation}
\label{Beqn}
B(p)=\sum_{l=2}^{\infty}z_{l}\sum_{k=1}^{l-1}pf^{k-1}(\fh+\sigma)^{l-k-1} .
\end{equation}

We then allow any number of nodes along the loop, though each time we
add a new node it creates a new edge of the loop.  Because of the
rotational symmetry, we divide by the number of nodes.  In addition,
there is a symmetry between the inside and the outside of the loop, leading to
a factor of $1/2$.  The total contribution thus becomes
\begin{equation}
\K_{1} = \frac{1}{2} \sum_{k=1}^{\infty} \frac{[y(A+B)]^{k}}{k} = -\frac{1}{2}\ln[1-y(A+B)] .
\end{equation}
Finally, to ensure that we have an odd number of odd nodes along the
loop, we set
\begin{equation}
\label{K1eqn}
K_{1} = \frac{\K_{1}(p=1)-\K_{1}(p=-1)}{2} .
\end{equation}

This function then generates all the diagrams with $2n$ channels.  We
can now choose any of the leaves to be labelled $i_1$, which fixes the
numbering of all other leaves: they are numbered in order along the
boundary walk.  The freedom of choosing one of the leaves gives a
factor of $2n$.  To get this factor we differentiate the result with
respect to $r$ and multiply by $r$ so that the power of $r$ still
counts the total number of channels.  Thus we obtain the generating
function
\begin{equation}
\label{FKeqn}
F = r\frac{\rmd K}{\rmd r} .
\end{equation}
%


For the transmission, using the semiclassical values of the 
variables in \eref{transsemivalues}, we find that the node 
contribution in \eref{Aeqn} becomes
\begin{equation}
A(p)=\frac{Nh(h-p-2)}{(1-h)^{2}} .
\end{equation}
As we do not allow the nodes to enter the leads (as the incoming and
outgoing channels are now in different leads) we also have $B=0$.
Note that the node contribution is given solely in terms of $h=\p\ph$
and the full contribution evaluates to
\begin{equation}
K_{1}=\frac{1}{4}\ln\left(\frac{1-h}{1+h}\right) .
\end{equation}
Putting in the correct explicit solution for $h$ from \eref{htranseqn} 
and transforming according to \eref{FKeqn}, we find the following 
generating function for the orthogonal correction to the moments of 
the transmission eigenvalues
\begin{equation}
\label{T1seqn}
T_{1}(s)=-\frac{\xi s}{(1-s)(1-s+4\xi s)} ,
\end{equation}
where we set $s=r^2$ to generate the moments as the $n$-th 
moment involves $2n$ channels.  This order correction was 
previously treated using a RMT diagrammatic expansion \cite{bb96}, 
and can be derived by performing an asymptotic expansion in inverse 
channel number of the RMT result for arbitrary channel number of \cite{sm10}.

For the reflection we have $\fh=f$ and the node contributions in
\eref{Aeqn} and \eref{Beqn} are
\begin{equation}
A(p)=\frac{Nf^2(f^2-p-2)}{(1-f^2)^{2}},\qquad B(p)=\frac{pN\zeta_{1}r^2}{(1-rf)^{2}} .
\end{equation}
Using relation \eref{freflsummed} we can rewrite $B(p)$ as
\begin{equation}
B(p) = \frac{pN}{\zeta_1} \frac{f^2}{(1-f^2)^2},
\end{equation}
so that for the generating function we find
\begin{equation}
\label{K1refleqn}
K_{1}=\frac{1}{4}\ln\left(\frac{\zeta_1+\zeta_2f^2}{\zeta_1-\zeta_2f^2}\right) 
=\frac{1}{4}\ln\left(\frac{f}{2r\zeta_1-f}\right),
\end{equation}
where for the last term we simplified the numerator and 
denominator inside the logarithm by only keeping the 
remainder after polynomial division with respect to 
the quadratic for $f$ in \eref{freflquad}.  
Putting in the explicit solution from \eref{freflquad} 
and following \eref{FKeqn} we obtain the rather simple 
generating function for the orthogonal correction to the 
moments of the reflection eigenvalues
\begin{equation}
\label{reflorthogen}
R_{1}(s)=\frac{\xi s}{(1-4\xi s)} .
\end{equation}
Note that this result only depends on 
$\xi=\zeta_{1}(1-\zeta_{1})=\zeta_{1}\zeta_{2}$, 
which is not so obvious from \eref{K1refleqn} and \eref{freflquad}.  
However, the relations in \eref{refltransrel} and the fact that the 
trace of the identity matrix, being the respective number of channels, 
is only leading order in inverse channel number means that the 
dependence only on $\xi$ of the subleading transmission 
moments \eref{T1seqn} must be mirrored in the reflection 
moments.  For the reflection into lead 2 we simply swap 
$\zeta_1$ and $\zeta_2$, which clearly does not affect 
this order correction.

\subsection{Unitary correction} \label{unit}

\begin{figure}
\center
\includegraphics[width=0.6\textwidth]{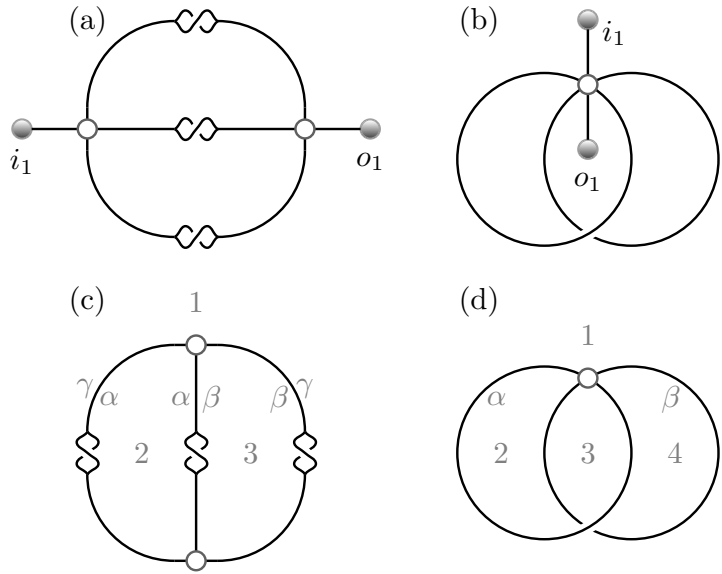}
\caption{\label{unittrees}The first subleading order semiclassical 
diagrams for systems without time reversal symmetry. We start with 
the trajectory pairs which contribute to the conductance in (a) and (b).  
Removing the channels and their links we obtain the base structures 
(c) and (d) for this case.}
\end{figure}

We can continue using the ideas above to treat higher order corrections.
In particular, for systems without time reversal symmetry the first
correction occurs at the second subleading order in inverse channel
number.  The semiclassical diagrams for the conductance are given, for
example, in \cite{mulleretal07} and can be represented as the graph
diagrams shown in \frefs{unittrees}(a) and (b).  We note that this
representation is not unique and is chosen for simplicity.  It is also
important to observe that, despite the twists, the corresponding
ribbon graphs are orientable, \ie have two surfaces (unlike the
M\"obius strip).  It can be shown this is true in general: diagrams
contributing to the unitary case are orientable.  Further, the
diagrams contributing at this order have genus 1, \ie embeddable on a
torus (but not a sphere).  This, too, can be shown to continue: the
contribution to the order $1/N^{(2g-1)}$ comes from diagrams of genus
$g$.

From the diagrams in \frefs{unittrees}(a) and (b) we can form the base structures
by removing the channels and their links, see \frefs{unittrees}(c) and
(d).  A similar restriction to the one above still holds when appending
subtrees to ensure that the resulting diagrams are permissible.
Namely, the total number of odd nodes and twists along every closed cycle in
the diagram has to be even.  We note that the definition of an odd
node depends on the cycle: the left node of \fref{unittrees}(a) is
odd with respect to the cycles formed from the top and bottom arcs and
is even relative to the cycle formed from the top arc and the middle
edge.  We remark that this rule was enforced for the M\"obius diagram
as well.

Finally, we need to discuss the symmetries of each base diagram.  The
generators of the symmetry group of base diagram \ref{unittrees}c
are the shift of the edge numbering and the reflection, giving a group
of size 6.  The generators for base diagram \ref{unittrees}d are
inside-outside mappings of the two edges, giving a group of size 4.

When we append subtrees along each edge of a diagram, whether we
append $f$ or $\fh$-type subtrees depends in a complicated way on the
types of subtrees appended along the other edges.  Therefore we
restrict ourselves to the simpler situation where $f=\fh$ and only
treat reflection quantities (with $\sigma=0$).  Along the links
connecting the nodes in the base diagrams we can append subtrees as
before, but because the rotational symmetry is now broken we no longer
divide by the number of nodes.  The edge contributions can therefore
be written as
\begin{equation}
E(p) = y\sum_{k=0}^{\infty} [y(A+B)]^{k} = \frac{y}{1-y(A+B)} ,
\end{equation}
where $A(p)$ and $B(p)$ are as in \eref{Aeqn} and \eref{Beqn} but 
with the simplification $f=\fh$ and $\sigma=0$.

We also need to append subtrees to the nodes of the base structures
and, finally, ensure that we have the correct number of objects (odd
nodes and twists) around each closed cycle.  To proceed, we number each
of the regions around the nodes and label the closed cycles with greek
letters as in \frefs{unittrees}(c) and (d).  We start with
\fref{unittrees}(c) and use powers of $p_{\alpha}$, $p_{\beta}$ and
$p_{\gamma}$ to count the number of objects along the respective
cycles.  At the top node we can add subtrees in any region we like as
long as we add an odd number in total to ensure that the top node
becomes a valid $l$-encounter.  An $l$-encounter involves $2l$
stretches and we have 3 stretches already from the base structure.  If
we place $k_{i}$ subtrees in each region $i$ and use the power of $q$
to count the total number of subtrees added, we can write the
contribution of the top node as
\begin{equation}
\V_{3\c}(q) = \sum_{k_1,k_2,k_3=0}^{\infty} x_{l}(qf)^{(k_1+k_2+k_3)}
p_{\alpha}^{k_{2}}p_{\beta}^{k_{3}}p_{\gamma}^{k_{1}},
\end{equation}
with $l=(k_1+k_2+k_3+3)/2$ and where the number 3 in the 
subscript refers to the fact that the node in the base 
diagram starts with 3 stretches while the `c' refers to 
its label in \fref{unittrees}.  Further, when we have an 
odd number of trees in each region, and when the odd numbered 
trees in each region are empty, then the top node can also 
enter the lead (since we are considering reflection quantities).  
If we define $k_i=2\k_i+1$ then we have $\k_i+1$ empty subtrees 
and $\k_i$ arbitrary subtrees in each region.  In total 
we would then add the contribution 
\begin{equation}
\V'_{3\c}(q) = qp_{\alpha}p_{\beta}p_{\gamma}\sum_{\k_1,\k_2,\k_3=0}^{\infty} z_{l}f^{(\k_1+\k_2+\k_3)} ,
\end{equation}
where $l=(\k_1+\k_2+\k_3+3)$ and we simplified the powers 
of $q$ and $p$ as in the end we are only interested in 
whether they are odd or even and they are all odd here.  
Finally to ensure that the total number of trees added is 
odd, we substitute
\begin{equation}
V_{3\c} = \frac{\V_{3\c}(q=1)-\V_{3c}(q=-1)}{2} + \V'_{3\c}(q=1).
\end{equation}

The complete diagram in \fref{unittrees}(c) is made up of two 
such nodes as well as three links.  Each of the links lies 
on two cycles so we can write the full contribution as
\begin{equation}
\K^{\U}_{2\c} = \frac{1}{6}E(p_{\alpha}p_{\beta})E(p_{\beta}p_{\gamma})E(p_{\gamma}p_{\alpha})(V_{3\c})^2
\end{equation}
where we divide by 6 to account for the symmetry of the base 
structure.  Here the number in the subscript now refers to the 
order of the contribution while the `U' in the superscript refers 
to the fact that these diagrams correspond to the unitary ensemble.  
Then to ensure that the number of objects along each cycle is even 
we simply average 
\begin{equation}
K^{\U}_{2\c} = \frac{\K^{\U}_{2\c}(p=1)+\K^{\U}_{2\c}(p=-1)}{2} ,
\end{equation}
for $p_{\alpha}$, $p_{\beta}$ and $p_{\gamma}$ in turn.

For the base structure in \fref{unittrees}(d) we now have a single 
node and four regions.  Region 3 lies inside both cycles and 
each cycle starts with a single object inside (and likewise 
outside) which are the stretches of the other cycle leaving 
or entering the node.  Treating the node as above, we obtain 
the contributions
\begin{equation}
\V_{4\d}(q) = \sum_{k_1,k_2,k_3,k_4=0}^{\infty} x_{l}(qf)^{(k_1+k_2+k_3+k_4)}
p_{\alpha}^{k_{2}}(p_{\alpha}p_{\beta})^{k_{3}}p_{\beta}^{k_{4}},
\end{equation}
with $l=(k_1+k_2+k_3+k_4+4)/2$ and
\begin{equation}
\V'_{4\d}(q) = \sum_{\k_1,\k_2,\k_3,\k_4=0}^{\infty} z_{l}f^{(\k_1+\k_2+\k_3+\k_4)} ,
\end{equation}
where $l=\k_1+\k_2+\k_3+\k_4+4$ and with an odd number of trees in
each region in this second case we are guaranteed to add an even number to each cycle and
an even number overall.  With four edges touching the node in the
base structure we need to add an even number of subtrees in total to
the node to make a valid $l$-encounter so the node contribution
reduces to
\begin{equation}
V_{4\d} = \frac{\V_{4\d}(q=1)+\V_{4\d}(q=-1)}{2} + \V'_{4\d}(q=1).
\end{equation}

Including the two edges we have a total contribution of
\begin{equation}
\K^{\U}_{2\d} = \frac{1}{4}p_{\alpha}p_{\beta}E(p_{\alpha})E(p_{\beta})V_{4\d} ,
\end{equation}
where we divide by 4 to account for the symmetry of the diagram 
and the $p_{\alpha}p_{\beta}$ accounts for the fact that the cycles 
each start with a single object (the original node, which is odd for 
both cycles).  We likewise take the average 
\begin{equation}
K^{\U}_{2\d} = \frac{\K^{\U}_{2\d}(p=1)+\K^{\U}_{2\d}(p=-1)}{2} ,
\end{equation}
for $p_{\alpha}$ and $p_{\beta}$ in turn.


When we put in the semiclassical substitutions from
\eref{reflsemivalues} the formulae above can be summed and simplified.
After applying the operator $r\frac{\rmd}{\rmd r}$ we find that first
correction for the reflection for the unitary case (adding the two
base cases) has the generating function
\begin{equation}
\label{reflunitgen}
NR^{\U}_{2}(s)=\frac{\xi^2 s^2(s-1)}{(1-4\xi s)^{\frac{5}{2}}} ,
\end{equation}

By restricting ourselves above to the situation where $f=\fh$ we are 
not able to obtain the transmission directly, but we can instead find 
the likely transmission generating function using \eref{refltransrel} 
that $t^{\dagger}t+r_{1}^{\dagger}r_{1}=I$
\begin{equation}
\label{transunitgen}
NT^{\U}_{2}(s)=-\frac{\xi^2 s^2}{(1-s)^{\frac{3}{2}}(1-s+4\xi s)^{\frac{5}{2}}} .
\end{equation}
The fact that we get such simple functions is a little surprising
especially as the result from each base case is notably more complex.
In fact this pattern can be seen to continue if we expand the RMT
result as in \ref{higherreflection}.  The generating function in
\eref{transunitgen} can also be obtained \cite{sm10} from their RMT result.

\subsection{Second orthogonal correction} \label{ortho2}

\begin{figure}
\center
\includegraphics[width=0.7\textwidth]{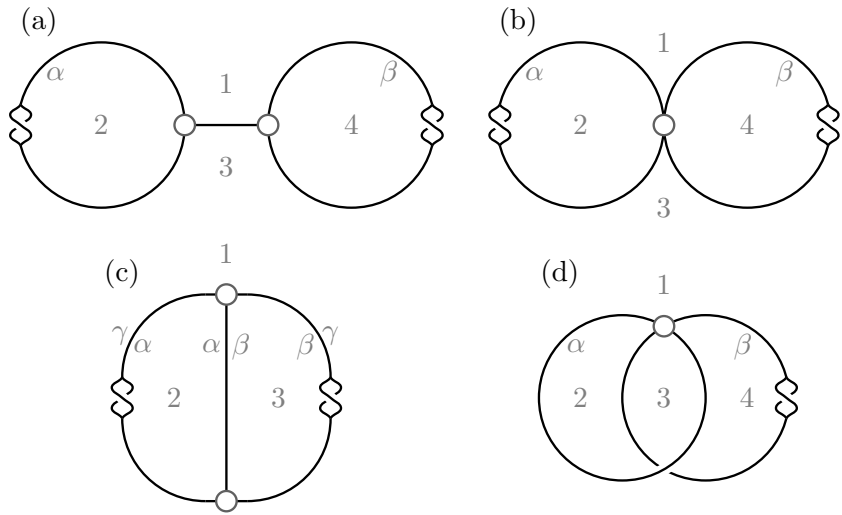}
\caption{\label{ortho2trees}The additional 4 base structures that 
exist for systems with time reversal symmetry at the second subleading 
order in inverse channel number.}
\end{figure}

When the system has time reversal symmetry, the edges and encounters
can again be traversed in different directions by the trajectory set
and their partners.  For the conductance there are 7 further
semiclassical diagrams at this order as depicted, for example in
\cite{mulleretal07}.  When we remove the starting and end links to
arrive at the base structures, we find that they reduce to the 4 base
cases depicted in \fref{ortho2trees}.

We note that the additional diagrams are non-orientable when viewed as
ribbon graphs.  Their groups of symmetry contain two elements each:
reflection for diagrams \ref{ortho2trees}(a)--(c) and inside-out flipping
of both edges simultaneously for diagram \ref{ortho2trees}(d).

\Frefs{ortho2trees}(c) and (d) are almost the same as \frefs{unittrees}(c) and (d), 
so we start with evaluating the contribution of \fref{ortho2trees}(a).  
Although region 1 and 3 are spatially connected they differ as to where 
we append subtrees at the nodes.  Starting with the node on the left we 
therefore get the contributions
\begin{equation}
\V_{3\a}(q) = \sum_{k_1,k_2,k_3=0}^{\infty} x_{l}(qf)^{(k_1+k_2+k_3)}p_{\alpha}^{k_{2}},
\end{equation}
with $l=(k_1+k_2+k_3+3)/2$ and
\begin{equation}
\V'_{3\a}(q) = qp_{\alpha}\sum_{\k_1,\k_2,\k_3=0}^{\infty} z_{l}f^{(\k_1+\k_2+\k_3)} .
\end{equation}
Again to ensure that an odd number of trees are appended, we substitute
\begin{equation}
V_{3\a} = \frac{\V_{3\a}(q=1)-\V_{3a}(q=-1)}{2} + \V'_{3\a}(q=1).
\end{equation}
For the node on the right we obtain the contribution $\hat{V}_{3\a}$ 
which is the same as $V_{3\a}$ but with $p_{\alpha}$ swapped with 
$p_{\beta}$ (and also $k_2$ with $k_4$).

Along with the two nodes in \fref{ortho2trees}(a) we have three links, 
two of which form cycles which already contain a single object (a twist). 
The total contribution is then
\begin{equation}
\K^{\O}_{2\a} = \frac{1}{2} p_{\alpha}p_{\beta}E(p_{\alpha})E(1)E(p_{\beta})V_{3\a}\hat{V}_{3\a} ,
\end{equation}
and to have an even number of objects along both cycles we average 
\begin{equation}
K^{\O}_{2\a} = \frac{\K^{\O}_{2\a}(p=1)+\K^{\O}_{2\a}(p=-1)}{2} ,
\end{equation}
for $p_{\alpha}$ and $p_{\beta}$ in turn.

The node in the base structure in \fref{ortho2trees}(b) provides 
the following contributions
\begin{equation}
\V_{4\b}(q) = \sum_{k_1,k_2,k_3,k_4=0}^{\infty} x_{l}(qf)^{(k_1+k_2+k_3+k_4)}p_{\alpha}^{k_{2}}p_{\beta}^{k_{4}},
\end{equation}
with $l=(k_1+k_2+k_3+k_4+4)/2$ and
\begin{equation}
\V'_{4\b}(q) = p_{\alpha}p_{\beta}\sum_{\k_1,\k_2,\k_3,\k_4=0}^{\infty} z_{l}f^{(\k_1+\k_2+\k_3+\k_4)} .
\end{equation}
The total number of trees added must be even, leading to
\begin{equation}
V_{4\b} = \frac{\V_{4\b}(q=1)+\V_{4\b}(q=-1)}{2} + \V'_{4\b}(q=1) ,
\end{equation}
while with the two cycles (which each start with a single object) 
we have a total contribution of
\begin{equation}
\K^{\O}_{2\b} = \frac12 p_{\alpha}p_{\beta}E(p_{\alpha})E(p_{\beta})V_{4\b} .
\end{equation}
As before we take the average 
\begin{equation}
K^{\O}_{2\b} = \frac{\K^{\O}_{2\b}(p=1)+\K^{\O}_{2\b}(p=-1)}{2} ,
\end{equation}
for $p_{\alpha}$ and $p_{\beta}$ in turn.

The difference of the structure in \fref{ortho2trees}(c) from 
that in \fref{unittrees}(c) is that now cycles $\alpha$ and $\beta$ 
start with an odd number of objects.  The contribution is then
\begin{equation}
\K^{\O}_{2\c} = \frac12 p_{\alpha}p_{\beta}E(p_{\alpha}p_{\beta})
E(p_{\beta}p_{\gamma})E(p_{\gamma}p_{\alpha})(V_{3\c})^2
\end{equation}
before averaging over the $p$'s in turn.  Only the $\alpha$ 
cycle in the structure in \fref{ortho2trees}(d) now starts with 
an odd number of objects so its contribution is
\begin{equation}
\K^{\O}_{2\d} = \frac12 p_{\alpha}E(p_{\alpha})E(p_{\beta})V_{4\d} ,
\end{equation}
and then we average over $p_{\alpha}$ and $p_{\beta}$ in turn.


Summing over the four new base cases (as well as the two that 
also exist without time reversal symmetry) we obtain the second 
subleading correction for the orthogonal case
\begin{equation}
\label{reflortho2gen}
NR^{\O}_{2}(s)=-\frac{\xi s\left[\xi s(3+s)+1-2s\right]}{(1-4\xi s)^{\frac{5}{2}}} .
\end{equation}
From this we can again find the likely generating function for the transmission
\begin{equation}
\label{transortho2gen}
NT^{\O}_{2}(s)=\frac{\xi s\left[\xi s(4s-3)+1-s^2\right]}{(1-s)^{\frac{3}{2}}(1-s+4\xi s)^{\frac{5}{2}}} .
\end{equation}
The moments generated by \eref{transortho2gen} can be proven \cite{sm10} to 
to agree with the moments obtained from an asymptotic expansion of their RMT 
result.  Semiclassically, time reversal symmetry allows more
possible diagrams, so the results here are somewhat more complicated
than the results \eref{reflunitgen} and \eref{transunitgen} for
systems without time reversal symmetry, but the RMT result \cite{sm10}
for the orthogonal case is notably more complex than for the unitary
case.  The results here are therefore useful in
simplifying asymptotic expansions of RMT moments.

\subsection{Leading order revisited} \label{leadingorder}

To obtain the leading order contributions \cite{bhn08}, the start of
the tree was fixed in the first incoming channel $i_1$ which allowed
the top node to possibly $i$-touch.  For example we simply place an
incoming channel on top of the tree in \fref{treecutting}(a). If the odd
numbered subtrees after the top node were empty then the top node could
$i$-touch the lead, leading to the generating function \cite{bk10}
\begin{equation}
F_{0}=c_{i}f+\sum_{l=2}^\infty z_{i,l} f^{l} +c_{o}\sigma+\sigma\sum_{l=2}^\infty z_{o,l} (\fh+\sigma)^{l-1} .
\label{Feqn}
\end{equation}
Here the terms involving $\sigma$ derive from the fact that the 
section of the tree above the top node is actually an empty $\fh$
tree.

However, using the ideas we developed for the subleading corrections
we can imagine a way of generating the leading order trees without
fixing any of the channels as a root.  As we shall see this is
particularly beneficial for calculating energy dependent correlation functions 
as in \srefs{andreev} and~\ref{wigner}.  To start, we view a single point as the base
structure for the leading order diagrams.  We therefore obtain the
leading order diagrams by joining subtrees to this point.  To create a
valid encounter we need to add $2l$ subtrees (with $l\geq2$), $l$ of
which are $f$-type subtrees starting from an incoming direction and
the other $l$ are $\fh$-type subtrees.  If all of the subtrees of a
particular type are empty then the node created can touch the lead.
We obtain the generating function
\begin{equation}
\label{Ktildeeqn}
\K_{0}=\sum_{l=2}^{\infty}\frac{x_{l}(f\fh)^{l}}{2l}
+\sum_{l=2}^{\infty}\frac{z_{i,l}f^{l}}{2l}
+\sum_{l=2}^{\infty}\frac{z_{o,l}(\fh+\sigma)^{l}}{2l} ,
\end{equation}
where we divide by $l$ because of the rotational symmetry and by a further 
factor of 2 because of the additional possibility of swapping the incoming 
and outgoing channels.  The last two terms in \eref{Ktildeeqn} represent 
moving the $l$-encounter (formed by appending the
subtrees to the starting point) into the incoming and outgoing channels.
Importantly we overcount the trees by the factor $V$ of their total
number of encounters because, for a given resulting tree, any of the nodes could
have been used as the base structure.

On the other hand we can also construct trees by joining \emph{two}
subtrees together, one $f$-type and one $\fh$-type.  After joining, the
new vertex of degree two gets absorbed into the edge.  A tree has
exactly $V-1$ internal edges, therefore there are $V-1$ ways to obtain a
given tree from joining two subtrees.  The joining operation gives the
contribution 
\begin{equation}
\label{Ktildeprimeeqn}
\K'_{0}=\frac{1}{2y}(f-yc_{o})(\fh-yc_{i}) - \frac{1}{2}c_{o}(yc_{i}+\sigma) ,
\end{equation}
where we divide by 2 because of the symmetry of swapping the incoming and outgoing 
channels.  In the first term in \eref{Ktildeprimeeqn} we subtract the empty tree 
from both $f$ quantities to ensure that they both include at least one node so that 
the edge formed is an internal one.  The last term in \eref{Ktildeprimeeqn} is then 
to ensure that the diagram made of a single diagonal link with no encounters ($V=0$) 
is included with the correct factor of $-1/2$.  Taking the difference between \eref{Ktildeeqn} and 
\eref{Ktildeprimeeqn} then means that we count each tree exactly once. 
Fittingly, for all the physical quantities we consider in this article 
\eref{Ktildeprimeeqn} equals minus the $l=1$ term in the sums in 
\eref{Ktildeeqn}.  This is natural because by joining two subtrees we 
essentially create a 1-encounter.  We simplify the difference to 
\begin{equation}
\label{K0eqn}
K_{0}=\K_{0}-\K'_{0} =\sum_{l=1}^{\infty}\frac{x_{l}(f\fh)^{l}}{2l}
+\sum_{l=1}^{\infty}\frac{z_{i,l}f^{l}}{2l}
+\sum_{l=1}^{\infty}\frac{z_{o,l}(\fh+\sigma)^{l}}{2l} .
\end{equation}

Now that no root is fixed we can make any channel to be the
first incoming channel and this generating function indeed misses a
factor $2n$ compared to the generating function $F$.  This turns out to
be very useful for the density of states of Andreev billiards in \sref{andreev} and to
recover $F$ we can use relation \eref{FKeqn}.


The generating function \eref{Feqn} for the reflection into lead 1 becomes
\begin{equation} 
\label{Frefleqn}
F_{0}=\frac{rN\zeta_{1}f}{1-rf} .
\end{equation}
Taking the solution of \eref{freflquad} or inverting \eref{Frefleqn} and 
substituting into \eref{freflquad} leads to the generating function for the reflection
\begin{equation}
\label{reflgeneqn}
\frac{R_{0}(s)}{N}=\frac{2\zeta_{1} s-1+\sqrt{1-4\xi s}}{2(1-s)} .
\end{equation}
For the reflection into lead 2 we swap $\zeta_1$ and $\zeta_2$.  
Note again that when we take the explicit solutions to any of the generating functions 
we chose the solution whose expansion in $r$ agrees with the semiclassical diagrams.

If, on the other hand, we start with \eref{K0eqn}, we obtain the generating 
function without the factor $2n$,
\begin{equation}
\label{Krefleqn}
\frac{K_{0}}{N}=\frac{1}{2}\ln(1-f^2)-\zeta_{1}\ln(1-rf) .
\end{equation}
To obtain the missing factor $2n$ we substitute the correct solution
of \eref{freflquad} into \eref{Krefleqn} and apply the operator $r\frac{\rmd}{\rmd r}$, 
as in \eref{FKeqn}.  Simplifying the result we recover
\eref{reflgeneqn}.

For the leading order transmission moments we also start with
\eref{Frefleqn} so with \eref{ftransfinaleqn} we obtain \cite{bhn08}
\begin{equation}
\label{transgeneqn}
\frac{T_{0}(s)}{N}=\frac{1}{2}\sqrt{1+\frac{4\xi s}{1-s}}-\frac{1}{2} .
\end{equation}
The integrated generating function is
\begin{equation}
\label{Ktranseqn}
\frac{K_{0}}{N}=\frac{1}{2}\ln(1-\p\ph)-\frac{\zeta_{1}}{2}\ln(1-r\p)
-\frac{\zeta_{2}}{2}\ln(1-r\ph) .
\end{equation}
As for the reflection, substituting the solutions of \eref{ftransfinaleqn} 
and transforming according to \eref{FKeqn} we recover \eref{transgeneqn}.

\addtocontents{toc}{\vspace{-1em}}
\section{Density of states of Andreev billiards} \label{andreev}

If we imagine merging the scattering leads and replacing them by a superconductor, 
then our chaotic cavity becomes an Andreev billiard.  Using the scattering 
approach \cite{beenakker05}, the density of states (normalised by its average) 
of such a billiard can be written as \cite{ihraetal01}
\begin{equation}
d(\epsilon)=1 + 2 \Imag \sum_{n=1}^{\infty} \frac{(-1)^{n}}{n} 
\frac{\partial C(\epsilon,n)}{\partial \epsilon} ,
\label{dossemieqn}
\end{equation}
in terms of energy-dependent correlation functions of the full scattering matrix
\begin{equation}
C(\epsilon,n)= \frac{1}{N}\Tr \left[S^{\dagger}\left(E-\frac{\epsilon\hbar}{2\tD}\right)
S\left(E+\frac{\epsilon\hbar}{2\tD}\right)\right]^n .
\end{equation}
Here the energy difference is in units of the Thouless energy
$E_{\mathrm{T}}=\hbar/2\tD$ (which depends on the average classical
dwell time $\tD$) and measured relative to the Fermi energy $E$.
Strictly speaking, for Andreev billiards we should use $S^{*}$ (matrix
$S$ with complex-conjugated entries) instead of the adjoint matrix
$S^{\dagger}$ but we will only consider systems with time reversal
symmetry where $S$ is symmetric.  With a superconductor at the lead,
each time the particle (electron or hole) hits a channel it is
retroreflected as the opposite particle (hole or electron) and
semiclassically (see \cite{kuipersetal10,kuipersetal10b} for fuller
details) we traverse the partner trajectories in the opposite
direction than for the reflection or transmission.  For the leading
order diagrams, this means that all the links (and encounters) are
traversed in opposite directions by electrons and holes, so that if we
break the time reversal symmetry, with a magnetic field say, then none
of these diagrams are possible any longer.  Interestingly, at
subleading order some diagrams are still allowed when the symmetry is
completely broken, as for example the coherent backscattering
contribution which comes from moving the node in \fref{boundarywalks}(e)
into the lead.

We can consider the generating function
\begin{equation}
  \label{eq:G_def}
  G(s)= \sum_{n=1}^{\infty} s^n C(\epsilon,n) ,
\end{equation}
which generates the required correlation functions.  We note that the 
definition of $G$ here is marginally different than in \cite{kuipersetal10,kuipersetal10b}.  
The semiclassical treatment there just requires us to make the substitutions 
\begin{eqnarray}
y&=&\frac{1}{N(1-a)},\quad x_{l}=-N(1-la), \nonumber \\
z_{i,l}&=&z_{o,l}=r^lN,\quad c_{o}=c_{i}=rN, \quad \sigma=0,
\label{energydepsubseqn}
\end{eqnarray}
where $a=\rmi\epsilon$.

\subsection{Subtrees} \label{andreevsubtrees}

As $\bz_{i}=\bz_{o}$ we have $f=\fh$ so \eref{eq:recur_f} becomes
\begin{equation}
  \label{fcorreqn}
f(1-a) = r - \sum_{l=2}^\infty(1-la)f^{2l-1} + r\sum_{l=2}^\infty r^{l-1}f^{l-1} ,
\end{equation}
or
\begin{equation}
  \label{fcorreqn2}
\frac{f(1-a-f^2)}{(1-f^2)^2}=\frac{r}{(1-rf)} ,
\end{equation}
which reduces to the cubic
\begin{equation}
  \label{fcorrcube}
  f^3-r(1+a)f^2-(1-a)f+r=0 .
\end{equation}

\subsection{Leading order} \label{andreevleadingorder}

The leading order in inverse channel number was obtained semiclassically 
in \cite{kuipersetal10,kuipersetal10b}, and for the energy dependent correlation functions we have 
\begin{equation} 
\label{Fcorreqn}
F_{0}=\frac{rNf}{1-rf} ,
\end{equation}
setting $G_{0}=F_{0}/N$, inverting \eref{Fcorreqn} and substituting 
into \eref{fcorrcube} we obtain the cubic
\begin{equation} 
\label{Gcorreqn}
(1-s)^2{G_{0}}^3+s(3s+a-3){G_{0}}^2+s(3s+a-1){G_{0}}+s^2=0 .
\end{equation}
However, for the density of states the correlation function \eref{K0eqn}
\begin{equation}
\label{Kcorreqn}
\frac{K_{0}}{N}=\frac{1}{2}\ln(1-f^2)+\frac{af^2}{2(1-f^2)}-\ln(1-rf) ,
\end{equation}
turns out to be more useful.  Indeed with $G=F/N$ and comparing \eref{FKeqn} with
\eref{eq:G_def}, we see that
\begin{equation}
  \label{eq:K_0_as_integrated_G}
  \frac{K_0}{N} = \sum_{n=1}^\infty \frac{s^n}{2n} C(\epsilon, n).
\end{equation}
Introducing
\begin{equation}
  \label{Hcorreqn}
  H(s)=\sum_{n=1}^{\infty}\frac{s^n}{n}\frac{\partial
    C(\epsilon,n)}{\partial a} = 2\frac{\partial}{\partial a} \frac{K_0}{N} ,
\end{equation}
this is precisely what is required for the density of states of Andreev
billiards in \eref{dossemieqn}, as by setting $s=-1$ we have
\begin{equation}
 \label{dossemieqn2}
d(\epsilon)=1+2\Real H(s=-1) .
\end{equation}
Performing the energy differential implicitly, we arrive at the cubic
\begin{equation}
\label{Hcubiceqn}
\fl a^2(1-s){H_{0}}^3+a\left[s(a-2)+2(1-a)\right]{H_{0}}^2
+\left[s(1-2a)-(1-a)^{2}\right]{H_{0}}+s=0 .
\end{equation}
Having the generating function $K$ therefore allows us easier access 
to the density of states than we had previously \cite{kuipersetal10,kuipersetal10b}. 
To make the connection to the RMT treatment, we set $s=-1$ and make the final substitution
\begin{equation}
 H_{0}(s=-1)=\frac{\left[\rmi W_{0}(\epsilon)-1\right]}{2}
\end{equation}
so that the leading order contribution to the density of states is 
$d_{0}(\epsilon)=-\Imag W_{0}(\epsilon)$ where $W_{0}$ satisfies the cubic
\begin{equation}
\epsilon^2{W_{0}}^3+4\epsilon {W_{0}}^2+(4+\epsilon^2)W_{0}+4\epsilon=0 ,
 \label{rmtsimplest}
\end{equation}
as found previously using RMT \cite{melsenetal96}.  The result can be written explicitly as
\begin{equation}
 d_{0}(\epsilon)=\Real \frac{\sqrt{3}}{6\epsilon}\left[Q_{0,+}(\epsilon)-Q_{0,-}(\epsilon) \right] ,
 \label{singledossol}
\end{equation}
where
\begin{equation}
Q_{0,\pm}(\epsilon)=\left[8-36\epsilon^2\pm3\epsilon\sqrt{3D}\right]^{\frac{1}{3}},
 \qquad D=\epsilon^4+44\epsilon^2-16 .
\end{equation}
Note that the density of states is only non-zero when the discriminant $D$ is positive, 
which occurs when $\epsilon>2\left(\frac{\sqrt{5}-1}{2}\right)^{\frac{5}{2}}$.

\subsection{First correction}\label{andreevone}

With the techniques in this article we can go beyond this, and RMT, and see what happens 
at the next two orders in inverse channel number.  For the first subleading order the 
even node contribution in \eref{Aeqn} becomes
\begin{equation}
A(p)=\frac{f^2(f^2-p-2)}{(1-f^2)^2}+\frac{af^2(f^4-3f^2+2p+4)}{(1-f^2)^3} ,
\end{equation}
while the odd node contribution in \eref{Beqn} is
\begin{equation}
B(p) = \frac{pr^2}{(1-rf)^2} = \frac{pf^2(1-a-f^2)^2}{(1-f^2)^4},
\end{equation}
which we rewrite using \eref{fcorreqn2}.  This simplifies the generating function 
\eref{K1eqn}, which reduces to
\begin{equation}
K_{1}=\frac{1}{4}\ln\left(\frac{(1+a)f^4-(2-a^2)f^2+1-a}{(1+a)f^4-(2+a^2)f^2+1-a}\right) .
\end{equation}

We provide the further generating functions in \ref{moregenfuncts},
but for the density of states we substitute $H_{1}(s=-1)=\rmi
W_{1}(\epsilon)/2$ in \eref{Hortheqn} and obtain the cubic
\begin{eqnarray}
 0&=&\frac{\epsilon^2}{4}D^2(NW_{1})^{3}+2\epsilon(11\epsilon^2-8)D(NW_{1})^{2} \nonumber \\
 & & {} +(\epsilon^6+528\epsilon^4-720\epsilon^2+256)NW_{1}+16\epsilon(3\epsilon^2+16) ,
\end{eqnarray}
so that the first correction to the density of states is given by
\begin{equation}
 d_{1}(\epsilon)=\frac{1}{N}\Real \frac{\sqrt{3}}{3\epsilon D}\left[Q_{1,+}(\epsilon)-Q_{1,-}(\epsilon) \right] ,
 \label{singledoscorrection}
\end{equation}
with
\begin{eqnarray}
Q_{1,\pm}(\epsilon)&=&\left[4096-46848\epsilon^2+103584\epsilon^4-3232\epsilon^6+126\epsilon^8 \right. \nonumber \\ 
& & \qquad \left. {} \pm3\epsilon(768-2928\epsilon^2+96\epsilon^4+\epsilon^6)\sqrt{3D}\right]^{\frac{1}{3}} .
\end{eqnarray}
As this involves the same discriminant as \eref{singledossol}, the correction 
to the density of states is also only non-zero when $\epsilon>2\left(\frac{\sqrt{5}-1}{2}\right)^{\frac{5}{2}}$, 
\ie it has the same gap as the leading order term.  The correction however is negative and has a singular peak 
from the discriminant in the denominator.

\subsection{Second correction}\label{andreevtwo}

Repeating this procedure for the six base cases that contribute at the second subleading order we find 
\begin{equation}
 d_{2}(\epsilon)=-\frac{1}{N^2}\Real \frac{\sqrt{3}}{3D^3}\left[Q_{2,+}(\epsilon)-Q_{2,-}(\epsilon) \right] ,
 \label{singledossecondcorrection}
\end{equation}
with
\begin{eqnarray}
\fl Q_{2,\pm}(\epsilon)&=&\left[2\epsilon D^{2}\left(16224878592+74096377856\epsilon^{2}
+153714421760\epsilon^{4}+86120095744\epsilon^{6} \right. \right. \nonumber \\
\fl & & \qquad {} +28154556672\epsilon^{8}+6522754176\epsilon^{10}+739116528\epsilon^{12}+
12120680\epsilon^{14} \nonumber \\ 
\fl & & \qquad \left. {} +122837\epsilon^{16}-5324\epsilon^{18}\right)
 \pm3D\left(3189506048-41603760128\epsilon^{2} \right. \nonumber \\ 
\fl & & \qquad {} -187618951168\epsilon^{4}-
192686981120\epsilon^{6} -237943482368\epsilon^{8}\nonumber \\
\fl & & \qquad {} -108211019264
\epsilon^{10}-13928492544\epsilon^{12} +1338160896\epsilon^
{14} \nonumber \\
\fl & & \qquad \left. \left. {} +274655180\epsilon^{16}-6774219\epsilon^{18}+
52756\epsilon^{20}\right)\sqrt{3D}\right]^{\frac{1}{3}} .
\end{eqnarray}

\begin{figure}
\center
\includegraphics[width=\textwidth]{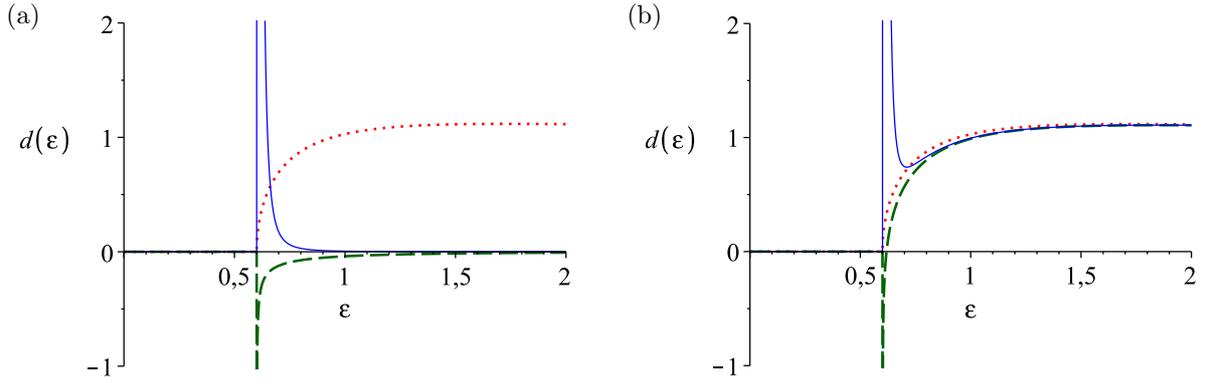}
\caption{\label{dosnew}(a) The leading order density of states (dotted) 
along with the first (dashed) and second (solid) correction for $N=25$.  
(b) The leading order density of states (dotted) with the first (dashed) 
and then the second (solid) correction added for $N=25$.}
\end{figure}

This correction is positive again and has a larger and steeper singular 
peak than the first order correction.  To illustrate this we plot the 
leading order result, as well as the two corrections in \fref{dosnew}(a) 
for $N=25$, while in \fref{dosnew}(b) we sequentially add the corrections 
to the leading order result, again for $N=25$.

The fact that the hard gap remains derives from the discriminant $D$ 
which is already present in $f$ (at $s=r^2=-1$) from \eref{fcorrcube}.  
We could then expect that the gap is robust against further higher 
order corrections, seeing as the expressions always involve $f$.  
Someway above the gap we see that the corrections (especially the second) 
make little difference but it is the region directly above the gap 
which is particularly interesting.  The expansion in inverse channel 
numbers is poorly (if at all) convergent but if the pattern of alternating 
singular peaks continues its sum (or rather the exact result for 
finite channel number) could take any value.  In particular the gap could widen.

Though treating the density of states of Andreev billiards
semiclassically basically just involves using different values for the
variables in the graphical recursions, on the RMT side it remains to
be seen how one could use recent advances, like the Selberg
integral approach, to proceed beyond the leading order in inverse
channel number.  For the leading order \cite{melsenetal96} a
diagrammatic expansion was performed, but a result for an
arbitrary number of channels would be especially welcome.  It would
determine whether the gap indeed persists and would clarify
what happens to the density of states just above the current gap.

\addtocontents{toc}{\vspace{-1em}}
\section{Moments of the Wigner delay times} \label{wigner}

Another quantity related to the energy dependent correlation functions
are the moments of the Wigner delay times.  To obtain them we define the
correlation function
\begin{equation}
\fl D(\epsilon,n)= \frac{1}{N}\Tr \left[S^{\dagger}
\left(E-\frac{\epsilon\hbar}{2\tD}\right)S\left(E+\frac{\epsilon\hbar}{2\tD}\right)
-S^{\dagger}\left(E\right)S\left(E\right)\right]^n ,
\label{Depseqn}
\end{equation}
where we subtract the identity matrix in the form of $S^{\dagger}S=I$.
 The corresponding generating function is
\begin{equation}
L(s)= \sum_{n=1}^{\infty} s^n D(\epsilon,n) .
\end{equation}
The moments of the Wigner-Smith matrix \cite{wigner55,smith60}
\begin{equation}
Q=\frac{\hbar}{\rmi}S^{\dagger}(E)\frac{\rmd S(E)}{\rmd E} ,
\end{equation}
are then given by \cite{bk10}
\begin{equation}
\label{motDepseqn}
m_{n}=\Tr \left[Q\right]^{n} = \frac{\tD^{n}}{\rmi^{n}n!}\frac{\rmd^{n}}{\rmd \epsilon^{n}} D(\epsilon,n)\Big\vert_{\epsilon=0} ,
\end{equation}
whose generating function we will denote
\begin{equation}
M(s)= \sum_{n=1}^{\infty} \frac{s^n}{\tD^{n}} m_n .
\end{equation}

The identity matrix in \eref{Depseqn} follows by removing the
$\epsilon$ dependence of the scattering matrices.  However, as the
identity matrix only has diagonal elements we identify these elements
as diagonal trajectory pairs that travel directly from incoming to
outgoing channels.  We considered these to be formed when we moved
encounters into the outgoing channels (equivalently we could use the
incoming channels instead) whenever we formed an empty $\fh$ subtree.
The empty subtree is included in the general $\fh$ contribution and we
included the contribution $\sigma$ to allow us to change the effective
value of $\fh$ in this situation.  The empty $\fh$ subtrees consist of
a single link and an incoming channel, producing the contribution
$yc_{i}=r/(1-a)$.  To mimic subtracting the identity matrix in
\eref{Depseqn} we simply take away the value of this empty
subtree at zero energy ($a=0$) by setting
\begin{equation}
\sigma=-r,
\end{equation}
along with the remaining semiclassical values in \eref{energydepsubseqn}.

\subsection{Subtrees} \label{wignersubtrees}

Including $\sigma$ breaks the symmetry of $f$ and $\fh$, so the subtree 
recursions \eref{eq:recur_f} and \eref{eq:recur_fhat} become 
\begin{equation}
  \label{motfeqns}
\fl \frac{f(1-a-f\fh)}{(1-f\fh)^2}=\frac{r}{(1+r^2-r\fh)} , 
\qquad \frac{\fh(1-a-f\fh)}{(1-f\fh)^2}=\frac{r}{(1-rf)} ,
\end{equation}
and we find \cite{bk10} that $f$ satisfies the following cubic
\begin{equation}
  \label{fmotcube}
 (1+r^2)^{2}f^3-r(1+r^2)(1+a)f^2-(1+r^2)(1-a)f+r=0,
\end{equation}
while $\fh$ is related by
\begin{equation}
\label{fhatmoteqn}
\fh =(1+r^2)f .
\end{equation}

\subsection{Leading order} \label{wignerleadingorder}

For the energy dependent correlation functions we have 
\begin{equation} 
\label{Fmoteqn}
F_{0}=\frac{rNf}{1-rf} - \frac{r^2N}{(1+r^2)(1-rf)} .
\end{equation}
Setting $L_{0}=F_{0}/N$, inverting \eref{Fmoteqn} and substituting 
into \eref{fcorrcube} we obtain the cubic
\begin{equation} 
\label{Lmoteqn}
(1+s){L_{0}}^3+as(1+s){L_{0}}^2+s(2as+a-1){L_{0}}+as^2=0 .
\end{equation}
For the $n$-th moment of the delay times we want the coefficient 
of the $n$-th power of $a$ which we can extract by transforming 
$s\to s/a$ and then setting $a=0$.  This leads to a quadratic 
and the leading order moment generating function \cite{bfb99,bk10}
\begin{equation}
\label{motlogeneqn}
M_{0}(s) = \frac{1-s-\sqrt{1-6s+s^2}}{2}
\end{equation}

Alternatively we can start with the generating function from \eref{K0eqn}
\begin{eqnarray}
\frac{K_{0}}{N}&=&\frac{1}{2}\ln(1-(1+r^2)f^2)+\frac{a(1+r^2)f^2}{2(1-(1+r^2)f^2)} \nonumber \\
& & -\frac{1}{2}\ln(1-rf)-\frac{1}{2}\ln\left[(1+r^2)(1-rf)\right] \label{Kmoteqn}
\end{eqnarray}
from which we can recover \eref{Lmoteqn} and hence \eref{motlogeneqn} 
by differentiating with respect to $r$, multiplying by $r$ and using 
the result for $\frac{\rmd f}{\rmd r}$ from differentiating \eref{fmotcube} 
implicitly.

\subsection{First orthogonal correction} \label{wignerone}

For the first orthogonal correction we evaluate the contributions 
of the even and odd nodes around the M\"obius strip
\begin{equation}
A(p)=\frac{h(h-p-2)}{(1-h)^2}+\frac{ah(h^2-3h+2p+4)}{(1-h)^3} ,
\end{equation}
and
\begin{equation}
B(p) = \frac{pr^2}{(1+r^2-r\fh)(1-rf)} = \frac{ph(1-a-h)^2}{(1-h)^4},
\end{equation}
which we again rewrite using \eref{motfeqns} and both contributions 
only depend on $h=f\fh$.  Putting these contributions into the 
generating function \eref{K1eqn} we again obtain
\begin{equation}
K_{1}=\frac{1}{4}\ln\left(\frac{(1+a)h^2-(2-a^2)h+1-a}{(1+a)h^2-(2+a^2)h+1-a}\right) ,
\end{equation}
as in \sref{andreevone} but with different values for $f$ and $\fh$ as given 
in \eref{fmotcube} and \eref{fhatmoteqn}.  Differentiating in line with \eref{FKeqn} 
and differentiating \eref{fmotcube} implicitly we arrive at the generating function, 
given as \eref{motortheqn} in \ref{moregenfuncts}, which generates the orthogonal 
correction to the correlation functions $D(\epsilon,n)$.  Finally by transforming 
$s\to s/a$ and setting $a=0$ we find the correction to the moments of the delay times to be
\begin{equation}
\label{motorthogeneqn}
NM_{1}(s) = \frac{1-3s-\sqrt{1-6s+s^2}}{2(1-6s+s^2)}.
\end{equation}

\subsection{Next corrections} \label{wignertwo}

Since we only treated reflection quantities where $f=\fh$ for the next
order corrections we cannot obtain the corresponding generating
functions of the moments of the delay times.  Instead we can generate the 
energy dependent correlation functions $C(\epsilon,n)$ by expanding the 
generating function $G_2(s)$ which can be found by treating the six base cases 
as in \sref{andreevtwo}.  Expanding to finite order, we can then obtain the functions
$D(\epsilon,n)$ using the relation
\begin{equation} 
  \label{DCreleqn}
  D(\epsilon,n) = \sum_{k=0}^{n}
  \left(-1\right)^{n-k} {n\choose k} 
  C(\epsilon,k) ,
\end{equation}
which follows from the binomial expansion of \eref{Depseqn} and where $C(\epsilon,0)=1$ has no 
subleading order contribution.  Doing this only for the two base structures which exist without 
time reversal symmetry and plugging the resultant $D(\epsilon,n)$ into 
\eref{motDepseqn} we obtain the moments for the unitary case to low order.  
We find that the generating function
\begin{equation}
\label{motunitgeneqn}
N^2M^{\U}_{2}(s) = \frac{2s^2}{(1-6s+s^2)^{\frac{5}{2}}} ,
\end{equation}
fits with these moments, while if we treat all six base structures, 
the generating function
\begin{equation}
\label{motortho2geneqn}
N^2M^{\O}_{2}(s) = \frac{s(s-3)}{\left(1-6s+s^2\right)^2}+\frac{3s(s-1)^2+2s^2}{\left(1-6s+s^2\right)^{\frac{5}{2}}} ,
\end{equation}
fits the low moments for systems with time reversal symmetry.

\addtocontents{toc}{\vspace{-1em}}
\section{Cross correlation of transport moments} \label{twotraces}

\begin{figure}
\center
\includegraphics[width=\textwidth]{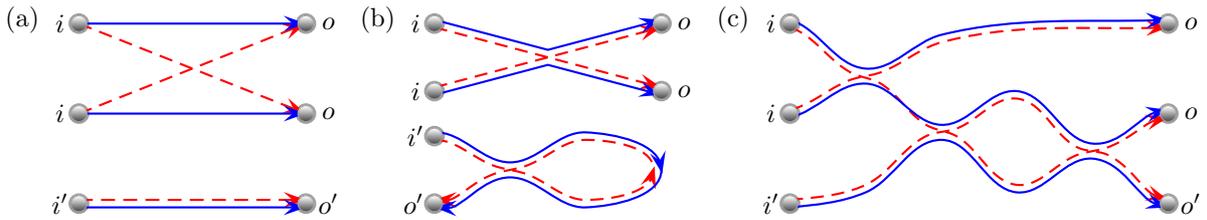}
\caption{\label{twotracediag}(a) With two traces, the semiclassical 
trajectories separate into two closed cycles.  When the two sets do not 
interact as in (b) we recreate terms from $\left\langle\Tr[X_{1}^{\dagger}X_{1}]^{n_1}\right\rangle
\left\langle\Tr[X_{2}^{\dagger}X_{2}]^{n_2}\right\rangle$, but when 
they do interact as in (c) further diagrams are possible as in \fref{twotracetree}.}
\end{figure}

Along with transport moments, we can also consider non-linear
statistics such as the cross correlation between transport moments, generated by
\begin{equation}
\label{Pdefeqn}
P_{[X_1,X_2]}(s_1,s_2) = \sum_{n_1,n_2=1}^{\infty}s_{1}^{n_{1}}s_{2}^{n_{2}}
\left\langle\Tr[X_{1}^{\dagger}X_{1}]^{n_1}\Tr[X_{2}^{\dagger}X_{2}]^{n_2}\right\rangle ,
\end{equation}
which involves two traces inside the energy average.  Semiclassically
we then have an expression involving two trajectory sets that form two
separate cycles, as in \fref{twotracediag}(a).  Of course when we look
for trajectory sets which lead to a small action difference we can
have independent diagrams for each set, as in \fref{twotracediag}(b).
However, these are included in the individual moments treated
previously, and when we remove them,
\begin{equation}
\fl \tilde{P}_{[X_1,X_2]}(s_1,s_2) = P_{[X_1,X_2]}(s_1,s_2) - 
\sum_{n_1,n_2=1}^{\infty}s_{1}^{n_{1}}s_{2}^{n_{2}}
\left\langle\Tr[X_{1}^{\dagger}X_{1}]^{n_1}\right\rangle
\left\langle\Tr[X_{2}^{\dagger}X_{2}]^{n_2}\right\rangle,
\end{equation}
we are left with trajectories that must interact, as in \fref{twotracediag}(c).
In RMT, this quantity is known as the `connected' part of the correlation
function.

It is interesting to consider the combinatorial interpretation of the
interacting sets of trajectories.  Denote the incoming channels belonging to
the first trace by $i_j$, $j=1,\ldots, n_1$ and the incoming channels
from the second trace by $i_j'$, $j=1,\ldots, n_2$.
Input channels are mapped onto output channels by trajectories from $X$
and also by trajectories from $X^\dagger$.  If we apply the first
mapping followed by the inverse of the second mapping, we end up with the
following transitions
\begin{equation}
  \label{eq:i_trans}
  i_1 \mapsto i_2, \ \ldots \ i_{n_1}\mapsto i_1,
  \qquad\mbox{and}\qquad
  i_1' \mapsto i_2', \ \ldots \ i_{n_2}'\mapsto i_1'.
\end{equation}
Thus the overall result is a permutation, written in the cycle
notation as $\pi=(i_1\ \ldots\ i_{n_1})(i_1'\ \ldots\ i_{n_2}')$.  We
can interpret each $l$-encounter as a cycle permuting $l$ labels of
the corresponding $X$ trajectories.  Then an entire leading order
diagram can be interpreted as a \emph{factorization} of $\pi$ into
smaller cycles (\cf \cite{bhn08b}).  In the language of
combinatorics, interacting trajectories correspond to a
\emph{transitive factorization}, leading order corresponds to the
\emph{minimality} condition and the fact that encounters on different
`branches' have no ordering imposed on them corresponds to counting
\emph{inequivalent factorizations} (up to a permutation of commuting
factors).  To summarize, the leading order interacting diagrams are in
one-to-one correspondence with the minimal transitive inequivalent
factorizations of a permutation into smaller cycles.  This question
has been studied combinatorially (for factorizations into
transpositions only) for a permutation consisting of two cycles in
\cite{GouJacLat_cjm01} and for three and more cycles (these correspond
to 3-point and higher cross correlations) in
\cite{Irv_thes04,BerIrv_prep}.  We note that the above combinatorial
questions and the evaluation of transport properties are not
completely equivalent problems.  To evaluate transport properties we
need to find additional information regarding the number of encounters
touching the lead.  On the other hand, we make substitutions
\eref{reflsemivalues} or \eref{transsemivalues} which significantly
simplify the results.

\subsection{Leading order}

\begin{figure}
\center
\includegraphics[width=0.7\textwidth]{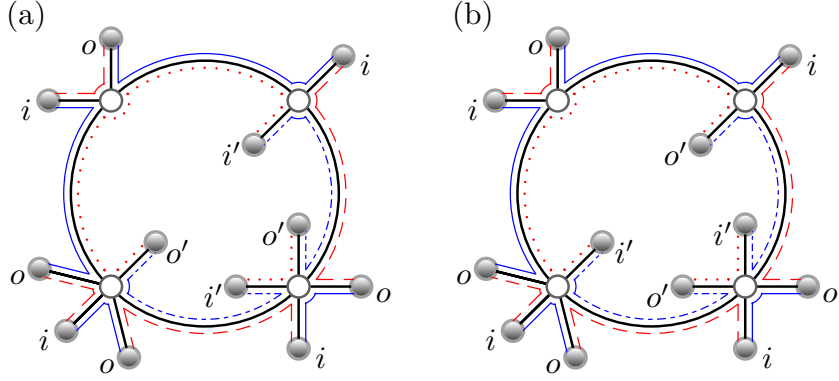}
\caption{\label{twotracetree}Example graphs, which contribute to the 
leading order in inverse channel number of $\tilde{P}_{[X_1,X_2]}(s_1,s_2)$, 
made out of a single loop with subtrees attached at nodes.  
For a small action difference, the (blue) solid or dashed dotted 
lines must partner (red) dashed or dotted lines on the other side 
of the loop.  (a) Without time reversal symmetry the odd nodes have 
either incoming or outgoing channels on the outer subtrees on each 
side of the loop. (b) With time reversal symmetry we may also swap 
the incoming and outgoing channels inside the loop.}
\end{figure}

We now proceed to expand the contributions of the interacting sets of
trajectories in inverse powers of the total channel number by
describing the corresponding graphical representations.  The base
diagram for the leading order term is just a single loop, like in
\fref{orthotree}(d) except with no twist.  Without the twist, the two
cycles of the permutation arise from the two walks on the inside and
outside of the loop, see \fref{twotracetree}(a).  One requirement, to
ensure a small action difference, is that the parts of the loop are
traversed on either side by parts of trajectories that contribute
actions with different signs in the semiclassical expression.  In
\fref{twotracetree} this means that the (blue) solid or dashed dotted
lines on either side of the loop must partner (red) dashed or dotted
lines on the other.  Without time reversal symmetry the parts of the
loop must additionally be traversed in the same direction by
trajectory stretches and their partners, so that at odd nodes (those
with an odd number of subtrees on each side of the loop) there is
unequal number of channels of a given type, as in
\fref{twotracetree}(a).  With time reversal symmetry, parts of the loop
may be traversed in any direction and we may also swap all the
incoming and outgoing directions on one side (the inside say) of the
loop as in \fref{twotracetree}(b).

With these restrictions we can start to append subtrees at nodes
around the loop.  We will use the tree function $f_1$ and generating
variable $\rho_1$ for the subtrees outside the loop and $f_2$ and
$\rho_2$ for those inside.  At each node we can either add an even or
odd number of subtrees on each side of the loop and we start with the
contribution when we add an even number
\begin{equation}
\label{Aeveneqn}
A^{\mathrm{even}}=\sum_{l=2}^{\infty}x_{l}\sum_{k=0}^{l-1}h_{1}^{k}h_{2}^{(l-1-k)} 
= \sum_{l=2}^{\infty}x_{l}\frac{h_1^l-h_2^l}{h_1-h_2},
\end{equation}
where $h_1=f_1\fh_1$.

For systems without time reversal symmetry, with an odd number of 
subtrees on each side (an odd node) we have two possibilities as 
in \fref{twotracetree}(a).  The subtrees in the odd positions on 
both sides all connect (first and last) to either incoming or 
outgoing channels.  With outgoing channels we have the contribution
\begin{equation}
\label{Aoddaoeqn}
\fl A^{\mathrm{odd}}_{\a,o}=\sum_{l=2}^{\infty}x_{l}\sum_{k=1}^{l-1}f_1h_{1}^{k-1}f_2h_{2}^{(l-1-k)} 
= \sum_{l=2}^{\infty}x_{l}f_1f_2\frac{h_1^{l-1}-h_2^{l-1}}{h_1-h_2},
\end{equation}
while with incoming channels we swap $f$ with $\fh$
\begin{equation}
\label{Aoddaieqn}
A^{\mathrm{odd}}_{\a,i}= \sum_{l=2}^{\infty}x_{l}\fh_1\fh_2\frac{h_1^{l-1}-h_2^{l-1}}{h_1-h_2}.
\end{equation}

As before it is possible that an odd node touches the lead when 
the odd positioned subtrees on both sides are empty.  Of course 
this also requires that the incoming or outgoing channels of the 
two quantities $X_1$ and $X_2$ originate or end in the same lead.  
When this is the case, we also have the following contribution
\begin{eqnarray}
\label{Baoeqn}
B_{\a,o} & = &\sum_{l=2}^{\infty}\frac{z_{o,l}}{r^l}
\sum_{k=1}^{l-1}\rho_1^{k}\fh_1^{k-1}\rho_2^{l-k}\fh_2^{(l-k-1)} \\ 
\nonumber &=& \sum_{l=2}^{\infty}\frac{z_{o,l}}{r^l}\rho_1\rho_2
\frac{(\rho_1\fh_1)^{l-1}-(\rho_2\fh_2)^{l-1}}{\rho_1\fh_1-\rho_2\fh_2} ,
\end{eqnarray}
where we needed to include the explicit $\rho_1$ and $\rho_2$ dependence of 
$\bz_{o}$ on the number of empty trees on each side of the loop. 
 With incoming channels instead we again swap $f$ and $\fh$
\begin{equation}
\label{Baieqn}
B_{\a,i} = \sum_{l=2}^{\infty}\frac{z_{i,l}}{r^l}\rho_1\rho_2
\frac{(\rho_1f_1)^{l-1}-(\rho_2f_2)^{l-1}}{\rho_1f_1-\rho_2f_2} .
\end{equation}

We allow an arbitrary number, $k$, of nodes along the loop, but the
total number of odd nodes must be even.  Taking into account
rotational symmetry, we get
\begin{equation}
\label{tkappaeqn}
\tilde{\kappa}_{1\a} = -\ln\left[1-yZ_{\a}(p)\right],
\end{equation}
where $Z_{\a}(p)$ is a contribution of one node.  The node can either be even or
odd of one of two types.  Since the number of odd $i$-nodes is equal
to the number of odd $o$-nodes, we can write $Z_{\a}(p)$ as
\begin{equation}
Z_{\a}(p) = A^{\mathrm{even}} + p \sqrt{(A^{\mathrm{odd}}_{\a,i}+B_{\a,i})(A^{\mathrm{odd}}_{\a,o}+B_{\a,o})}.
\end{equation}
To ensure that we indeed have an even number of odd nodes, we set
\begin{equation}
\label{kappa1eqn}
\kappa_{1} = \frac{\tilde{\kappa}_{1}(p=1)+\tilde{\kappa}_{1}(p=-1)}{2} ,
\end{equation}
leading to
\begin{equation}
  \label{kappa1eqn2a}
  \fl \kappa_{1\a} = -\frac12 \ln\left[(1-yA^{\mathrm{even}})^{2} 
    - y^2(A^{\mathrm{odd}}_{\a,i}+B_{\a,i})(A^{\mathrm{odd}}_{\a,o}+B_{\a,o})\right].
\end{equation}

For the additional contribution, in the case of systems with time reversal symmetry,
from the diagrams like \fref{twotracetree}(b) we swap the incoming and
outgoing channels inside the loop so that we now have the
contributions
\begin{equation}
\label{Aoddbeqns}
\fl A^{\mathrm{odd}}_{\b,o} = \sum_{l=2}^{\infty}x_{l}\fh_1f_2\frac{h_1^{l-1}-h_2^{l-1}}{h_1-h_2}, 
\qquad A^{\mathrm{odd}}_{\b,i}= \sum_{l=2}^{\infty}x_{l}f_1\fh_2\frac{h_1^{l-1}-h_2^{l-1}}{h_1-h_2}, 
\end{equation}
and
\begin{eqnarray}
\label{Bbeqns}
B_{\b,o} &=& \sum_{l=2}^{\infty}\frac{z_{o,l}}{r^l}\rho_1\rho_2
\frac{(\rho_1f_1)^{l-1}-(\rho_2\fh_2)^{l-1}}{\rho_1f_1-\rho_2\fh_2} , \\
\nonumber B_{\b,i} &=& \sum_{l=2}^{\infty}\frac{z_{i,l}}{r^l}\rho_1\rho_2
\frac{(\rho_1\fh_1)^{l-1}-(\rho_2f_2)^{l-1}}{\rho_1\fh_1-\rho_2f_2} ,
\end{eqnarray}
and correspondingly
\begin{equation}
\label{kappa1eqn2b}
\fl \kappa_{1\b} = -\frac12 \ln\left[(1-yA^{\mathrm{even}})^{2} 
- y^2(A^{\mathrm{odd}}_{\b,i}+B_{\mathrm{b},i})(A^{\mathrm{odd}}_{\b,o}+B_{\b,o})\right].
\end{equation}

There is also an additional freedom of placing the label
$i_1$ on any leaf outside, giving a factor of $2n_1$.  Once $i_1$ has
been placed the type (in- or out-) of the leaves inside is fixed and
the freedom of placing the label $i_1'$ inside produces only a factor
of $n_2$.  We obtain these factors by differentiating with respect to
the variables $\rho_1$ and $\rho_2$,
\begin{equation}
\label{gammakappaeqn}
\Gamma=\frac{\rho_1\rho_2}{2}\frac{\partial^{2} \kappa}{\partial \rho_1 \partial \rho_2} .
\end{equation}
We further note that the differentiation ensures that there are at least two channels both
inside and outside the loop.

Using the appropriate semiclassical values of the variables $\bx$, $\bz$ 
and $y$ as well as the corresponding subtree contributions, we find 
the following generating functions
\begin{equation}
  \label{eq:1corr_rr}
\fl \tilde{P}_{[r_1,r_1],1}^{\U}(s_1,s_2) = 
\tilde{P}_{[r_1,r_2],1}^{\U}(s_1,s_2) = 
\frac {s_1s_2}{2\left(s_1-s_2\right)^{2}}
\left[\frac{1-2\xi\left(s_1+s_2\right) }{\sqrt {1-4\xi s_1}\sqrt {1-4\xi s_2}} - 1\right] ,
\end{equation}
with $s_1=\rho_1^2$ and $s_2=\rho_2^2$ and twice this result for the
orthogonal case with time reversal symmetry.  Even though for the
autocorrelation $\tilde{P}_{[r_1,r_1],1}^{\U}$ we can always move the
odd nodes into the lead while for the cross correlation
$\tilde{P}_{[r_1,r_2],1}^{\U}$ we can not, this is somehow compensated
for by the different subtree contributions and both give the same
result.

For the transmission autocorrelation we can move the odd nodes into 
the lead only for the unitary diagrams, leading to the generating function
\begin{equation}
  \label{eq:1corr_tt}
\fl \tilde{P}_{[t,t],1}^{\U}(s_1,s_2) 
= \frac {s_1s_2}{2\left(s_1-s_2\right)^{2}}\left[\frac{1+\left(2\xi-1\right) 
 \left(s_1+s_2\right) + \left(1-4\xi \right)s_1s_2 }{(1-s_1)
\sqrt {1+\frac{4\xi s_1}{1-s_1}}(1-s_2)\sqrt {1+\frac{4\xi s_2}{1-s_2}}}-1\right],
\end{equation}
and we still obtain twice this for the orthogonal result.  Finally 
for the cross correlation between the reflection and transmission we have
\begin{equation}
  \label{eq:1corr_rt}
\fl \tilde{P}_{[r_1,t],1}^{\U}(s_1,s_2) 
= \frac {s_1s_2}{2\left(s_1-s_1s_2+s_2\right)^{2}}
\left[1-\frac{1-s_2+2\xi \left(s_2-s_1+s_1s_2\right)}
{\sqrt {1-4\xi s_1}(1-s_2)\sqrt {1+\frac{4\xi s_2}{1-s_2}}}\right],
\end{equation}
and twice this for the orthogonal case.  Note that the above results
remain unchanged if we swap $r_1$ and $r_2$ as this just means
swapping $\zeta_1$ and $\zeta_2$ in the semiclassical contributions
which does not change $\xi$.  From these results we can obtain the
corresponding $P_{[X_1,X_2]}$ up to the first subleading order by
including the first three orders in inverse channel number of the
moments corresponding to $X_1$ multiplied by the moments corresponding
to $X_2$.  If we also include $n=0$ terms (which are just the number
of channels in the respective lead) with those moments then we obtain
the $n_1=0$ and $n_2=0$ terms in \eref{Pdefeqn}.  This then allows us
to check that expansions of the various transport correlation
functions indeed fulfil the unitarity conditions in
\eref{refltransrel}.  Note that if we assume \emph{a priori} that the
unitarity is preserved by the semiclassical
approximation \eref{scatmateqn}, any one of equations
\eref{eq:1corr_rr}--\eref{eq:1corr_rt} implies all others.

\subsection{Subleading correction} \label{twotracesnextcorrection}

\begin{figure}
\center
\includegraphics[width=0.7\textwidth]{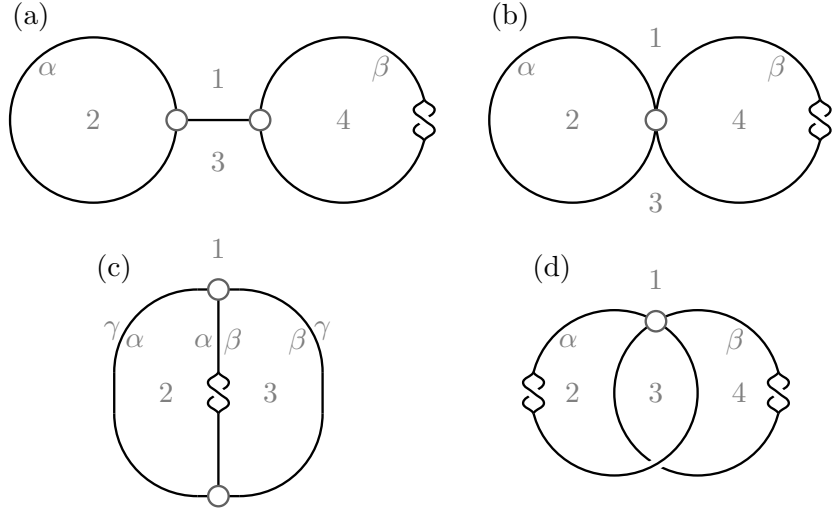}
\caption{\label{ortho2corr}The base structures which break into 
two cycles (for systems with time reversal symmetry) at the second 
subleading order in inverse channel number.}
\end{figure}

We can continue this process and look at the base structures like in 
\frefs{unittrees} and \ref{ortho2trees} but which separate into two cycles.  
In fact the possibilities are almost the same as in \fref{ortho2trees} 
but with one twist more or fewer as depicted in \fref{ortho2corr}.  
These can also exist only for systems with time reversal symmetry 
and we can treat them in a similar way as before, but with the modifications above.

Again the types of subtrees at each node depends on the nodes 
elsewhere in the diagram, so we restrict out attention to the 
simpler case of the reflection where $f=\fh$.  Because of 
\eref{refltransrel}, we have
\begin{equation}
\Tr[r_{1}^{\dagger}r_{1}]^{n} - N_1 =  \sum_{k=1}^n (-1)^k  {n\choose k}  
\Tr[t^{\dagger}t]^{k}= \Tr[r_{2}^{\dagger}r_{2}]^{n} - N_2 ,
\end{equation}
so that $\tilde{P}_{[r_1,r_1]} = \tilde{P}_{[r_1,r_2]}$ where the 
reflection autocorrelation is equal to the reflection cross correlation.  
However for the cross correlation, as the channels of the two 
reflections are in different leads, the nodes which lie on both 
cycles can not enter the lead, and this further simplifies the 
calculation.   The edges which travel through both cycles then provide factors
\begin{equation}
\eta(p) = \frac{1}{1-(A^{\mathrm{even}}+pA^{\mathrm{odd}})} ,
\end{equation}
while the edges which only pass through one cycle provide factors 
$E(p)$ as before, but with the semiclassical values corresponding 
to the reflection into lead 1 or lead 2 as appropriate.  We will 
denote this correspondence by a subscript in the following.

The treatment of the diagrams is very similar to that in \sref{ortho2}
so we merely highlight the steps here.  But first we discuss the
symmetry factors.  Because of time-reversal symmetry, we can put the
first incoming channels on both faces on any leaf, leading to the
differential operator $\rho_1\rho_2\frac{\partial^{2} \kappa}{\partial
\rho_1 \partial \rho_2}$.  Unlike the diagrams in \fref{twotracetree}, the
`inside' and `outside' faces are, in general, not related by symmetry,
and we should consider both putting $f_1$-trees on the `outside' and
on the `inside'.  For brevity we will only list the former
contributions.  Finally, the symmetry groups of diagrams
\ref{ortho2corr}(c) and \ref{ortho2corr}(d) have order $2$ and $4$
correspondingly and we will divide their contributions by the
appropriate factor.

Starting with the diagram in \fref{ortho2corr}(a), for the node on
the left, which cannot enter the lead, we have
\begin{equation}
\v_{3\a}(q) = - \sum_{k_1,k_2,k_3=0}^{\infty} (qf_1)^{(k_1+k_3)}(qf_2)^{k_2}p_{\alpha}^{k_{2}},
\end{equation}
with an odd number of trees appended
\begin{equation}
\nu_{3\a} = \frac{\v_{3\a}(q=1)-\v_{3a}(q=-1)}{2} .
\end{equation}
For the node on the right we have $\hat{V}_{3\a}$ as before but 
with semiclassical values corresponding to the reflection into 
lead 1.  To ensure a valid semiclassical diagram we still need 
each cycle to contain an even number of objects, so we have
\begin{equation}
\ka^{\O}_{2\a} = p_{\beta}\eta(p_{\alpha})E_{1}(1)E_{1}(p_{\beta})\nu_{3\a}\hat{V}_{3\a}.
\end{equation}
This is then averaged 
\begin{equation}
\kappa^{\O}_{2\a} = \frac{\ka^{\O}_{2\a}(p=1)+\ka^{\O}_{2\a}(p=-1)}{2} ,
\end{equation}
for $p_{\alpha}$ and $p_{\beta}$ in turn.  Finally, we add the
contribution where we place $f_1$ subtrees along the inside and $f_2$
subtrees along the outside.  For the reflection cross correlation this
reduces to swapping $\rho_1$ with $\rho_2$ and swapping $\zeta_1$ with
$\zeta_2=1-\zeta_1$.

The node in \fref{ortho2corr}(b) gives
\begin{equation}
\v_{4\b}(q) = - \sum_{k_1,k_2,k_3,k_4=0}^{\infty} (qf_1)^{(k_1+k_3+k_4)}(qf_2)^{k_2}p_{\alpha}^{k_{2}}p_{\beta}^{k_{4}},
\end{equation}
with an even number of subtrees in total
\begin{equation}
\nu_{4\b} = \frac{\v_{4\b}(q=1)+\v_{4\b}(q=-1)}{2} ,
\end{equation}
The total contribution is then
\begin{equation}
\ka^{\O}_{2\b} = p_{\beta}\eta(p_{\alpha})E_{1}(p_{\beta})\nu_{4\b} .
\end{equation}
averaged over $p_{\alpha}$ and $p_{\beta}$ in turn and we again add
the result where we swap the trees on the inside and the outside.

For the structure in \fref{ortho2corr}(c) the nodes provide 
\begin{equation}
\v_{3\c}(q) = -\sum_{k_1,k_2,k_3=0}^{\infty} (qf_1)^{k_1}(qf_2)^{(k_2+k_3)}
p_{\alpha}^{k_{2}}p_{\beta}^{k_{3}}p_{\gamma}^{k_{1}},
\end{equation}
with  
\begin{equation}
\nu_{3\c} = \frac{\v_{3\c}(q=1)-\v_{3c}(q=-1)}{2} ,
\end{equation}
so that the contribution is then
\begin{equation}
  \k^{\O}_{2\c} = \frac{1}{2}
  p_{\alpha}p_{\beta} E_{2}(p_{\alpha}p_{\beta}) \eta(p_{\beta}
  p_{\gamma}) \eta(p_{\gamma} p_{\alpha}) (\nu_{3\c})^2
\end{equation}
before averaging over the $p$'s in turn.  Likewise we include 
the contribution where we swap the subtrees on the inside with 
those on the outside.

Finally the node in \fref{ortho2corr}(d) provides
\begin{equation}
\fl \v_{4\d}(q) = -\sum_{k_1,k_2,k_3,k_4=0}^{\infty} (qf_{1})^{(k_1+k_3)}(qf_{2})^{(k_2+k_4)}
p_{\alpha}^{k_{2}}(p_{\alpha}p_{\beta})^{k_{3}}p_{\beta}^{k_{4}},
\end{equation}
with 
\begin{equation}
\nu_{4\d} = \frac{\v_{4\d}(q=1)+\v_{4\d}(q=-1)}{2}.
\end{equation}
The total contribution is then
\begin{equation}
\ka^{\O}_{2\d} = \frac{1}{4} \eta(p_{\alpha}) \eta(p_{\beta}) \nu_{4\d} ,
\end{equation}
averaged over $p_{\alpha}$ and $p_{\beta}$ in turn.  We also
include the contribution where we swap the subtrees on the inside 
and outside. 

Summing the four diagrams, we eventually arrive at the generating function
\begin{eqnarray}
\fl N\tilde{P}_{[r_1,r_2],2}^{\O}(s_1,s_2)&=&\left[s_1+s_2-2s_1s_2 -2\xi
\left(4s_1^2+s_2^2+3s_1s_2-s_1^3-5s_1^2s_2-2s_1s_2^2\right)\right.  \nonumber \\
\fl & & \qquad \left. {} + 8s_1^2\xi^2\left(s_1+3s_2-3s_1s_2-s_2^2\right)\right]\nonumber \\
\fl & & \times  \frac{s_1s_2}{(s_1-s_2)^{3}(1-4\xi s_1)^2\sqrt{1-4\xi s_2}} + \left(s_1 \leftrightarrow s_2\right),
\end{eqnarray}
where $\left(s_1 \leftrightarrow s_2\right)$ means we add the result with $s_1$ and $s_2$ swapped.

We could check that the results in this section all agree with the first four moments
calculated from the arbitrary channel RMT results of \cite{ssw08}.

\addtocontents{toc}{\vspace{-1em}}
\section{Conclusions and discussion} \label{concs}

We described a method for the semiclassical calculation of the
expansion of several transport statistics asymptotically in the
inverse channel number $1/N$.  The calculation is performed by
grafting trees onto the base structures with a low number of cycles,
and relies on the fact that attaching trees does not change the order
in inverse channel number.  Instead, the trees add more incoming and
outgoing channels and so increase the order of the moment.  With
graphical recursions, this allows us to generate all the moments at a
given order in inverse channel number, which we performed up to the
third order.  The terms we considered suggest the following
observations about the ribbon graphs that arise as the contributing
diagrams
\begin{itemize}
\item absence of time reversal symmetry results in graphs being
  orientable; both orientable and non-orientable graphs contribute to
  the calculation with time reversal symmetry,
\item the order (in $1/N$) of a contribution is reflected in the genus
  of the corresponding graph,
\item linear moments (with one trace) result in graphs with one face,
  while non-linear moments with $m$ traces will require considering
  graphs with $m$ faces.
\end{itemize}
The above general observations suggest that a complete expansion in
$1/N$ should be both feasible and interesting to specialists working
in algebra and combinatorics.

We find that the semiclassical contribution of individual base
diagrams depends significantly on the global structure of the diagram.
This is in contrast to the expansions of the first two moments
performed in \cite{heusleretal06,braunetal06,mulleretal07}, where the
contribution factorized into a product over the vertices of the
diagram and the problem was thus reduced to a combinatorial
enumeration.  The latter was achieved through finding recursion
relations which connect different diagrams, and similar ideas could
well be useful for the base structures we need for all moments.  To
illustrate the scale of the problem of going to higher order in $1/N$,
for the unitary case there are 1848 base diagrams at the next
contributing order.  As they can involve more than two nodes, they
can no longer be derived by cleaning the corresponding semiclassical
conductance diagrams as was the case for the orders treated
in this article.  However, in the end all these diagrams would
probably lead to the generating functions \eref{unitreflorder3}
and \eref{unittransorder3}, highlighting the scale of the
simplifications that take place.

Our results fully agree with the predictions of RMT theory (as far as
those are available \cite{sm10}), and importantly are given in terms 
of very simple generating functions. 
This would suggest that extending the types of asymptotic analyses of 
\cite{novaes07,carreetal10,krattenhaler10} beyond the leading order, 
(as is currently being performed \cite{sm10}), one 
could also expect to see simplifications of the RMT results.   For the unitary 
case, where several different formulae are known for the moments of the 
transmission eigenvalues \cite{vv08,novaes08,sm10,lv11}, this analysis 
and the semiclassical endpoints could shed light on the combinatorial 
relationships between the different approaches.
Because of the connection between RMT and weakly disordered systems,
we can expect that our results also apply to such systems.  Likewise, with
the close correspondence between semiclassical and disorder diagrams \cite{sla98},
one might hope to find similar graphical recursions in a perturbative 
expansion of the appropriate nonlinear $\sigma$ model.

If we were to consider non-linear statistics
involving three traces (with their mean parts removed), the leading
contribution would come from the diagrams like in \frefs{ortho2trees}
and \ref{ortho2corr} which split into three cycles, \ie diagrams (a)--(c)
without any twists.  Considering quantities with $m$ traces we would
then need to treat base diagrams related to those which contribute to
order $1/N^{(m-2)}$ (and higher) for the linear statistics, and so we
immediately run into the considerations and difficulties described
above.  Curiously though the moments of the conductance and the shot
noise themselves can be efficiently treated using RMT 
\cite{novaes08,ok08,ok09,kss09}.  Semiclassically, the $m$-th moment of these quantities
corresponds to having exactly 2 or 4 channels along each of the $m$ cycles,
and the RMT results might then provide a pathway for generating 
such semiclassical diagrams.  This could in turn be useful for generating 
and treating the corresponding base diagrams for the linear statistics.

The methods described in this article were also used to treat the
density of states of Andreev billiards.  Replacing the normal
conducting leads of the chaotic cavity by a superconductor produces
strong effects like the complete suppression of the density of states
around the Fermi energy
\cite{melsenetal96,kuipersetal10,kuipersetal10b}.  Being interested in
the density of states one must evaluate moments of all orders,
something that our methods are particularly geared towards.  Going beyond
leading order in inverse channel number we could show that this gap
persists for the next two orders, and that the behaviour of the
density of states slightly above the gap is not determined by just
these terms in the expansion.  Because of the superconductor, one not
only needs to know all the moments but also all the higher orders in
inverse channel number.  A result for arbitrary channel number would
therefore be particularly welcome for such systems, which leads to the
question of how to adapt the recent RMT advances to tackle this
problem.  Similarly, chaotic cavities with additional superconductors
attached (Andreev dots) also exhibit significant effects due to the
presence of the superconductors and also require one to be able to
treat what would correspond to all the moments of usual transport
quantities.  For example, at leading order in inverse channel number,
the conductance through a normal chaotic cavity requires just the
diagonal pair of trajectories, while for the conductance through an
Andreev dot one needs full tree recursions \cite{ekr10}.  The
treatment is actually similar to the edges in the base diagrams here,
but with the added ingredient of having two (or more) different species
of subtree.  One can then see that treating the transport moments of
Andreev dots requires an extra layer of complexity compared to normal
chaotic cavities.

The results in this article are all for the case in which the leads are
perfectly coupled to the chaotic cavity, rather than for the more general
and experimentally relevant case of non-ideal coupling.  
This is typically modelled by introducing tunnel
barriers into the leads with some probability to
backscatter when entering (or leaving) the cavity.
Semiclassically, along with affecting the contributions of the channels 
and modifying the survival probability and hence contributions of the links and the
correlated trajectory stretches inside the encounters, the main change is that
a wealth of new diagrams become possible \cite{whitney07}.  Specifically,
encounters may now partially touch the leads and have some of their links
backreflected at the tunnel barrier while the rest tunnel through to enter or exit the
system.  In principle, these possibilities would become extra terms 
in the tree and graphical recursions in this article, but so far these types of diagrams
have only been treated semiclassically for the lowest moments \cite{whitney07,kuipers09}.
However, from a RMT viewpoint at leading order in inverse channel number
and with the same tunnelling probability for each channel, the non-ideal
contacts just increase the order of the generating functions by one, 
for example for both the density of states of Andreev billiards \cite{melsenetal96}
and the moments of the Wigner delay times \cite{sss01}.

Finally one can wonder whether the effect of the Ehrenfest time can 
be incorporated into the graphical recursions developed here.  For the 
leading order in inverse channel number, the effect could be included 
\cite{wkr10} in the tree recursions.  First, the trees are related 
to each other through a continuous deformation, for example by 
giving the nodes a certain size (actually, the Ehrenfest time itself) and 
allowing them to slide into each other. Second, this is then partitioned 
in a particular way so that one can extract the Ehrenfest time dependence efficiently.  
Each partition and hence the sum of all diagrams leads to the same 
simple Ehrenfest time dependence at leading order in inverse channel 
number \cite{wkr10} and this pattern and treatment seems to also hold 
at the first subleading order \cite{waltner10}.  Whether this continues to 
higher order and nonlinear statistics, which start to include the 
complications of periodic orbit encounters, is an intriguing question. 

\ack{The authors would like to thank Daniel Waltner for useful discussions 
and Klaus Richter for further discussions and hospitality during a visit of 
one of us (GB) to Regensburg University, kindly supported by the
Vielberth foundation.  We would also like to thank Nick Simm 
and Francesco Mezzadri for discussing and sharing their RMT results 
and we gratefully acknowledge the Alexander von Humboldt Foundation 
(JK) and the National Science Foundation under Grants DMS-0604859
and DMS-0907968 (GB) for funding.}

\appendix

\addtocontents{toc}{\vspace{-1em}}
\section[]{Unitary reflection and transmission} \label{higherreflection}

Using the RMT result from \cite{novaes08} we can compute the moments 
up to finite order and expand them in powers of the inverse channel 
number.  Looking at the patterns for the reflection in \eref{reflgeneqn},
\eref{reflorthogen} and \eref{reflunitgen}, we can expect that each 
order in the inverse channel number just increases the powers in the 
denominators (the square root comes simply from the subtrees).  
In fact we find that the first several subleading orders can be written as
\begin{equation}
N^{2k-1}R_{2k}(s)=\frac{\xi^2 s^2(s-1)}{(1-4\xi s)^{\frac{6k-1}{2}}}\cdot
 \boldsymbol{\chi}_{2k-1}X_{2k-1}\boldsymbol{S}_{2k-1}^{\mathrm{T}}
\end{equation}
where $\boldsymbol{\chi}_{m}$ is the vector 
$(1,\xi s,\xi^2 s^2,\ldots,\xi^{m-1}s^{m-1})$, $X_{m}$ is an $m\times m$ matrix and 
$\boldsymbol{S}_{m}$ is the row vector $(1,s,s^2,\ldots,s^{m-1})$.  The first few values
of $X_m$ are
\begin{equation}
X_1 = (1)
\end{equation}
\begin{equation}
\label{unitreflorder3}
X_3 = \left(\begin{array}{ccc}
1 & -8 & 8 \\
20 & -20 & -8 \\
9 & -2 & 9
\end{array}\right)
\end{equation}
\begin{equation}
X_5 = \left(\begin{array}{ccccc}
1 & -40 & 220 & -360 & 180 \\
136 & -1240 & 2480 & -1360 & -32 \\
1770 & -5700 & 4890 & -1392 & 528 \\
3080 & -4760 & 1736 & 408 & -720 \\
450 & -360 & 76 & -360 & 450
\end{array}\right)
\end{equation}
\begin{equation*} 
\fl X_7 = \left(\small{\begin{array}{ccccccc}
1 & - 168 & 2688 & -13104 & 26712 & -24192 & 8064 \\ 
636 & -19068 & 125832 & -313824 & 342048 & -151200 & 15552 \\ 
34659 & -401142 & 1357755 & -1917888 & 1210608 & -326496 & 42744 \\ 
398328 & -2303784 & 4455864 & -3638040 & 1157712 & 5952 & - 77312 \\ 
1152438 & -3903144 & 4477620 & -1989144 & 322326 & -186288 & 130032 \\ 
766584 & -1652952 & 1107456 & -238704 & - 35400 & 172872 & -126000 \\
55125 & -78750 & 28539 & - 5732 & 28539 & -78750 & 55125
\end{array}}\right)
\end{equation*}
%
%

Similarly we can write the transmission as
\begin{equation}
\fl N^{2k-1}T_{2k}(s)=-\frac{\xi^2 s^2}{(1-s)^{\frac{2k+1}{2}}(1-s+4\xi s)^{\frac{6k-1}{2}}}\cdot 
\boldsymbol{\upsilon}_{2k-1}Y_{2k-1}\boldsymbol{S}_{2k-1}^{\mathrm{T}}
\end{equation}
where $\boldsymbol{\upsilon}_{m}$ is the vector 
$((1-s)^{m-1},\xi s(1-s)^{m-2},\xi^2 s^2(1-s)^{m-3},\ldots,\xi^{m-1}s^{m-1})$
and $\boldsymbol{S}_{m}$ is the row vector $(1,s,s^2,\ldots,s^{m-1})$. 
The first few values of the matrix $Y_m$ are
\begin{equation}
Y_1 = (1)
\end{equation}
\begin{equation}
\label{unittransorder3}
Y_3 = \left(\begin{array}{ccc}
1 & 6 & 1 \\
-20 & 20 & 8 \\
9 & -16 & 16
\end{array}\right)
\end{equation}
\begin{equation}
Y_5 = \left(\begin{array}{ccccc}
1 & 36 & 106 & 36 & 1 \\
-136 & - 696 & 424 & 424 & 16 \\
1770 & - 1380 & - 1590 & 1632 & 96 \\
-3080 & 7560 & - 5936 & 1920 & 256 \\
450 & - 1440 & 1696 & - 512 & 256
\end{array}\right)
\end{equation}
\begin{equation*} 
\fl Y_7 = \left(\small{\begin{array}{ccccccc}
1 & 162 & 1863 & 4012 & 1863 & 162 & 1 \\ 
-636 & - 15252 & - 40032 & 11544 & 25572 & 3228 & 24 \\ 
34659 & 193188 & - 128070 & - 194892 & 111939 & 25680 & 240 \\
-398328 & 86184 & 1088136 & - 885864 & 84144 & 101760 & 1280 \\ 
1152438 & - 3011484 & 2248470 & 61344 & - 524256 & 199680 & 3840 \\ 
-766584 & 2946552 & - 4341456 & 2993280 & - 862464 & 150528 & 6144 \\ 
55125 & - 252000 & 461664 & - 423424 & 221952 & - 12288 & 4096
\end{array}}\right)
\end{equation*}
%
%

Similar patterns hold for the moments of the delay times 
for the unitary case, and the results are actually simpler 
than for the transmission and reflection since there is one 
parameter fewer.  The likely generating functions can be 
found by expanding the RMT result of \cite{sm10} and fitting 
to the behaviour of \eref{motlogeneqn} and \eref{motunitgeneqn}.

\addtocontents{toc}{\vspace{-1em}}
\section[]{Further generating functions} \label{moregenfuncts}

For the energy dependent correlation functions, we find 
that the generating function $G_{1}=F_{1}/N$ satisfies the cubic
\begin{eqnarray}
\fl 0 &=& \left[4(1-a)^3+s(a^4-20a^2-8)+4s(a+1)^3\right]^{2}(NG_{1})^{3} \nonumber \\
\fl & & {} + 2\left[(a-1)^3+s^2(a+1)^3 \right]\left[4(1-a)^3+s(a^4-20a^2-8)+4s^2(a+1)^3\right](NG_{1})^{2} \nonumber \\
\fl & & {} + \left[(a-1)^6-4s(a+1)(a-1)^4+3s^2(a-1)(a+1)(a^4-8a^2-2)\right. \nonumber \\
\fl & & {} \qquad \left. +4s^3(a-1)(a+1)^4+s^4(a+1)^6\right]NG_{1} \nonumber \\
\fl & & {} + a^2s\left[(a-1)^3+s^2(a+1)^3\right] , \label{Gortheqn}
\end{eqnarray}
while the energy differentiated generating function instead satisfies
\begin{eqnarray}
\fl 0 &=& a^2\left[4(1-a)^3+s(a^4-20a^2-8)+4s^2(a+1)^3\right]^{2}(NH_{1})^{3} \nonumber \\
\fl & & {} + 2a\left[(4-a)(a-1)^2-2s(5a^2+4)+s^2(a+4)(a+1)^2 \right]\nonumber \\
\fl & & {} \qquad \times \left[4(1-a)^3+s(a^4-20a^2-8)+4s^2(a+1)^3\right](NH_{1})^{2} \nonumber \\
\fl & & {} + \left[(a-4)^2(a-1)^4+4s(a-1)^2(6a^3-21a^2+4a-16)\right. \nonumber \\
\fl & & {} \qquad +s^2(-3a^6+156a^4+102a^2+96) \nonumber \\
\fl & & {} \qquad \left. -4s^3(a+1)^2(6a^3-21a^2+4a-16)+s^4(a+4)^2(a+1)^4\right]NH_{1} \nonumber \\
\fl & & {} + as(s-1)\left[(a+2)(a-4)^2+s(a-2)(a+4)^2\right] . \label{Hortheqn}
\end{eqnarray}

For the moments of the delay times we have the first subleading order generating function 
\begin{eqnarray}
\fl 0 &=& (s+1)^2\left[4(1-a)^3+sa(a^3-8a^2+4a-24)+s^2a^2(a^2+4)\right]^{2}(NL_{1})^{3} \nonumber \\
\fl & & {} + 2(s+1)\left[(a-1)^3+2s(a-1)^3+2s^2a(a^2+3)\right] \nonumber \\
\fl && {} \qquad \times \left[4(1-a)^3+sa(a^3-8a^2+4a-24)+s^2a^2(a^2+4)\right](NL_{1})^{2} \nonumber \\
\fl & & {} + \left[(a-1)^6+4sa(a-3)(a-1)^4+3s^2a^2(a-1)(3a^3-13a^2+20a-28)\right. \nonumber \\
\fl & & {} \qquad \left. +2s^3a^2(a-1)(5a^3-11a^2+16a-32)^4+s^4a^2(5a^4+27a^2+32)\right]NL_{1} \nonumber \\
\fl & & {} + sa^2\left[(a-1)^3+2s(a-1)^3+2s^2a(a^2+3)\right] . \label{motortheqn}
\end{eqnarray}

\end{document}